\documentclass[superscriptaddress,
 reprint,
 amssymb, amsmath,
 aip, cha
]{revtex4-1}
\usepackage[utf8]{inputenc}
\usepackage[T1]{fontenc}
\usepackage{amsmath}
\usepackage{amsfonts}
\usepackage{amssymb}
\usepackage{graphicx}
\usepackage{bm}
\usepackage{grffile} %Allows figure file names with periods
\usepackage{xr} %use for supplementary material
\usepackage[dvipsnames]{xcolor}
\usepackage{mathtools}

\graphicspath{{figures/}}

\begin{document}

%\maketitle

\title{Model reduction for the collective dynamics of globally coupled oscillators: From finite networks to the thermodynamic limit}

\author{Lachlan~D. Smith}
 \email{lachlan.smith@sydney.edu.au}
 \affiliation{\mbox{School of Mathematics and Statistics, The University of Sydney, Sydney, NSW 2006, Australia}}
\author{Georg~A. Gottwald}
 \email{georg.gottwald@sydney.edu.au}
 \affiliation{\mbox{School of Mathematics and Statistics, The University of Sydney, Sydney, NSW 2006, Australia}}

\date{\today}

\begin{abstract}
Model reduction techniques have been widely used to study the collective behavior of globally coupled oscillators. However, most approaches assume that there are infinitely many oscillators. Here we propose a new ansatz, based on the collective coordinate approach, that reproduces the collective dynamics of the Kuramoto model for finite networks to high accuracy, yields the same bifurcation structure in the thermodynamic limit of infinitely many oscillators as previous approaches, and additionally captures the dynamics of the order parameter in the thermodynamic limit, including critical slowing down that results from a cascade of saddle-node bifurcations.
\end{abstract}

\maketitle

\begin{quotation}
Model reduction methods reduce the dynamics of high-dimensional complex systems to a small number of active degrees of freedom, which enables theoretical and analytical understanding of observed phenomena. Here we expand on the recently introduced collective coordinate approach to study globally coupled oscillators by treating the order parameter explicitly as a collective coordinate and by introducing a new ansatz function. We achieve a model reduction which accurately captures the macroscopic dynamics of finite populations of oscillators, and also recovers well-known analytical results in the thermodynamic limit of infinitely many oscillators which were previously derived using self-consistency relations. Our approach enables deeper analytical insight into the dynamics of the system. For instance, the transition from global synchronization to partial synchronization, and then to incoherence, occurs for finite networks as a cascade of saddle-node bifurcations. This is reflected in the thermodynamic limit by a critical slowing down of the macroscopic dynamics.
\end{quotation}

\section{Introduction}

Many natural phenomena and industry applications can be modeled as networks of coupled oscillators, including firefly flashing \cite{MirolloStrogatz90}, neuron firing \cite{SheebaEtAl08, BhowmikShanahan12}, and power grid dynamics \cite{FilatrellaEtAl08}. A common phenomenon in networks of coupled oscillators is synchronization. Model reduction techniques aim to understand and quantify this low-dimensional emergent macroscopic dynamics. For a recent review of model reduction approaches see \citet{BickEtAl20}. For the Kuramoto model \cite{Kuramoto84, Strogatz00, PikovskyEtAl01, AcebronEtAl05, OsipovEtAl07, ArenasEtAl08, DorflerBullo14, RodriguesEtAl16}, which is widely used to model networks of coupled oscillators, Ott and Antonsen \cite{OttAntonsen08, OttAntonsen09} introduced a method that describes its low-dimensional dynamics by deriving a closed set of equations for mean-field variables restricted to an ansatz manifold. Many studies have since applied and generalized the Ott-Antonsen approach to describe low-dimensional phenomena such as chimera states \cite{PanaggioAbrams15, Laing09, Laing09_2}, cluster synchronization from higher order coupling \cite{SkardalEtAl11} or symplectic coupling \cite{SkardalArenas19}, chaotic intercluster dynamics \cite{BickEtAl18}, and hysteretic synchronization \cite{PazoErnest09}. While the Ott-Antonsen approach is exact under the assumptions of infinitely many oscillators and smooth frequency distributions, it cannot describe the collective behavior in real-world networks which are finite in size. In particular, the Ott-Antonsen approach cannot describe dynamical phenomena which are entirely determined by finite size effects, such as stochastic drift of the synchronized cluster in the stochastic Kuramoto model \cite{Lucon15, BertiniEtAl10, BertiniEtAl14, Gottwald17}. Another commonly used model reduction approach is Watanabe-Strogatz theory \cite{WatanabeStrogatz93} which yields an exact system of ordinary differential equations for a small number of macroscopic parameters, and is not restricted to the thermodynamic limit of infinitely many oscillators. However, the Watanabe-Strogatz approach only applies to populations of identical oscillators \cite{RosenblumPikovsky15}. The Ott-Antonsen and the Watanbe-Strogatz approaches can be connected in the thermodynamic limit \cite{PikovskyRosenblum11}.

Recently a new approach based on \textit{collective coordinates} was developed which is not restricted to the thermodynamic limit of infinitely many oscillators or to identical oscillators and which accurately describes the macroscopic dynamics of the Kuramoto model\cite{Gottwald15, Gottwald17, HancockGottwald18, SmithGottwald19, YueEtAl20}. We improve here the original collective coordinate framework by considering an improved ansatz function describing the shape of the synchronized cluster. We treat the order parameter as the collective coordinate, yielding evolution equations for its dynamics along a judiciously chosen ansatz manifold. We will show that both the previous ansatz function (which is based on a linearization) and the improved ansatz function quantitatively capture the collective dynamics for small finite populations of oscillators, accurately capturing finite size effects. The improved ansatz function yields a significant improvement compared to the previous ansatz function. Moreover, we will show that the improved ansatz describes the collective behavior of coupled oscillators across the whole range from finite networks to the thermodynamic limit. We will show that the new collective coordinate ansatz yields identical bifurcation structure as the Ott-Antonsen ansatz in the thermodynamic limit of infinitely many oscillators, recovering well-known conditions for partial synchronization. Unlike the Ott-Antonsen approach, the collective coordinate approach is also applicable to non-analytic natural frequency distributions, such as uniform distributions, and captures critical slowing down of the order parameter that results from a cascade of saddle-node bifurcations.

The paper is organized as follows. In Section~\ref{sec:coll_coord} the collective coordinate framework for model reduction is described and the new ansatz is presented. In Section~\ref{sec:CC-vs-OA} numerical and analytical results are presented for finite networks with several natural frequency distributions. In Section~\ref{sec:SN_bifurcation} we show that for finite networks the transition from synchronization to incoherence occurs as a cascade of saddle-node bifurcations. In Section~\ref{sec:therm_lim} the thermodynamic limit is studied. Section~\ref{sec:conclusions} summarizes the results.

\section{Collective coordinate reduction for finite networks} \label{sec:coll_coord}

For a network of $N$ coupled oscillators, each with phase $\phi_i$, the Kuramoto model \cite{Kuramoto84} with all-to-all coupling is given by
\begin{equation} \label{eq:full_KM}
\dot{\phi_i} = \omega_i + \frac{K}{N} \sum_{j=1}^N  \sin(\phi_j - \phi_i),
\end{equation}
where $K$ is the coupling strength and the natural frequencies $\omega_i$ have distribution $g(\omega)$. Without loss of generality we assume here that $g(\omega)$ has zero mean, as can be achieved by moving into a co-rotating reference frame. While not a necessary assumption for the collective coordinate framework, here we will consider examples where $g(\omega)$ is unimodal and symmetric.

The general method of collective coordinates is to assume an ansatz $\hat{\bm{\phi}}$ for the synchronized state, i.e. $\phi_i \approx \hat{\phi}_i(\alpha;\omega_i)$ for $i\in \mathcal{C}$, where $\mathcal{C}$ is the set of oscillators that partake in the synchronized dynamics. The collective coordinate $\alpha(t)$ controls the shape of the synchronized state. One then performs a Galerkin approximation of the Kuramoto model (\ref{eq:full_KM}) with the ansatz function. The error incurred by this ansatz is given by substituting the ansatz into the Kuramoto model (\ref{eq:full_KM}),
\begin{equation} \nonumber
\mathcal{E}_i = \dot{\alpha} \frac{d\hat{\phi}_i}{d\alpha} - \omega_i - \frac{K}{N} \sum_{j\in \mathcal{C}} \sin(\hat{\phi}_j - \hat{\phi}_i),
\end{equation}
for $i \in \mathcal{C}$. We ignore non-entrained ``rogue'' oscillators with $i\notin \mathcal{C}$ that do not partake in the collective synchronized dynamics. For symmetric frequency distributions, these rogue oscillators have no effect on the synchronized cluster in the thermodynamic limit, since rogue oscillators with positive frequencies cancel out corresponding rogue oscillators with negative frequencies. For finite networks with symmetric frequency distributions the effect of the rogue oscillators can also be assumed to be negligible \cite{YueEtAl20}, since the time-average of the fast rogue dynamics cancels to zero. Since we are assuming a solution to the Kuramoto model (\ref{eq:full_KM}) of the form $\bm{\phi} = \hat{\bm{\phi}}(\alpha)$, the error vector $\bm{\mathcal{E}}$ is minimized provided that it is orthogonal to the tangent space of the synchronization manifold spanned by $\frac{d\hat{\bm{\phi}}}{d\alpha}$. The condition 
\begin{equation}
\left\langle \bm{\mathcal{E}}, \frac{d\hat{\bm{\phi}}}{d\alpha}\right\rangle = 0,
\end{equation}
where $\langle\text{-},\text{-}\rangle$ denotes the Euclidean scalar product, then yields the evolution equation for the collective coordinate
\begin{equation}  \label{eq:CC_general_form}
\dot{\alpha} = \frac{1}{||\frac{d\hat{\bm{\phi}}}{d\alpha} ||^2}\left( \left\langle\bm{\omega} , \frac{d\hat{\bm{\phi}}}{d\alpha} \right\rangle + \frac{K}{N} \sum_{i,j\in \mathcal{C}} \frac{d\hat{\phi_i}}{d\alpha} \sin(\hat{\phi}_j - \hat{\phi}_i) \right).
\end{equation}
A stable stationary point $\alpha^\star$ of (\ref{eq:CC_general_form}) corresponds to a synchronized state $\phi_i = \hat{\phi}_i(\alpha^\star)$, for $i \in \mathcal{C}$. Under the hypothesis that all oscillators that can synchronize will synchronize, the set $\mathcal{C}$ is defined as the maximal set of oscillators such that stationary points $\alpha^\star$ of (\ref{eq:CC_general_form}) exist. The identification of $\mathcal{C}$ is discussed in more detail in Section~\ref{sec:cluster}. 

\begin{figure}[tbp]
\centering
\includegraphics[width=\columnwidth]{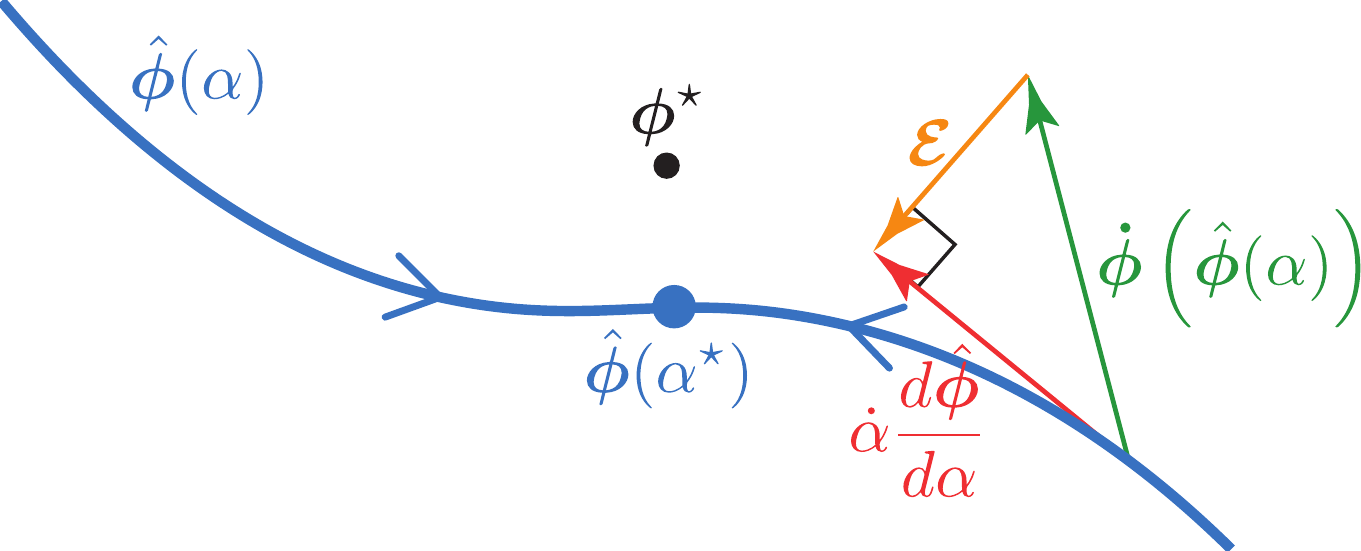}
\caption{The collective coordinate method describes the evolution of the Kuramoto model projected orthogonally onto the ansatz manifold $\hat{\bm{\phi}}(\alpha) \in \mathbb{R}^{|\mathcal{C}|}$ (blue).}
\label{fig:CC_schematic}
\end{figure}

The collective coordinate method is illustrated diagrammatically in Fig.~\ref{fig:CC_schematic}, where $\hat{\bm{\phi}}(\alpha)$ is the one-dimensional ansatz manifold in $\mathbb{R}^{|\mathcal{C}|}$, with $|\mathcal{C}|$ denoting the cardinality of $\mathcal{C}$. The collective coordinate method describes the evolution of the Kuramoto model projected orthogonally onto the ansatz manifold, with non-entrained rogue oscillators ignored, i.e.,
\begin{equation} \label{eq:CC_projection_form}
\Pi_{\frac{d\hat{\bm{\phi}}}{d\alpha}} \dot{\bm{\phi}} = \dot{\alpha} \frac{d\hat{\bm{\phi}}}{d\alpha},
\end{equation}
where $\dot{\bm{\phi}}$ is the dynamics of the full Kuramoto model (\ref{eq:full_KM}) and $\Pi_{\frac{d\hat{\bm{\phi}}}{d\alpha}}$ denotes orthogonal projection onto the tangent vector $\frac{d\hat{\bm{\phi}}}{d\alpha}$. The temporal evolution equation of the collective coordinate (\ref{eq:CC_general_form}) thus describes the dynamics restricted to the ansatz manifold. The full Kuramoto model converges to the stationary point $\bm{\phi}^\star \in \mathbb{R}^{|\mathcal{C}|}$ (given as a time-average when rogues are included), and the collective coordinate model converges to $\hat{\bm{\phi}}(\alpha^\star)$ which lies on the ansatz manifold $\hat{\bm{\phi}}(\alpha)$.

The collective coordinate framework requires to specify the shape function $\hat{\bm{\phi}}(\alpha ;\bm{\omega})$, and the choice of the collective coordinate $\alpha$ which parameterizes the ansatz manifold $\hat{\bm{\phi}}(\alpha ;\bm{\omega})$. To motivate the choice of $\hat{\bm{\phi}}$, we introduce the complex order parameter 
\begin{equation} \nonumber
z(t)=r(t) e^{i\psi(t)} = \frac{1}{N}\sum_j e^{i \phi_j}.
\end{equation}
The Kuramoto model (\ref{eq:full_KM}) can be rewritten as a mean-field equation
\begin{equation} \label{eq:KM_mean_field}
\dot{\phi_i} = \omega_i + K r \sin(\psi - \phi_i),
\end{equation}
where we have assumed that the mean natural frequency is zero \cite{Kuramoto84}. Similarly, we can assume that $\psi=0$ \footnote{This is a standard procedure \cite{Kuramoto84, Strogatz00}. The coordinate transformation is $\phi(t) \to \phi(t) - \Omega t - \psi$, where $\Omega$ is the mean frequency of the cluster.}. In the synchronized state, oscillators in $\mathcal{C}$ become approximately stationary, so that
\begin{equation} \label{eq:mean_field_solution}
\phi_i \approx  \arcsin\left( \frac{\omega_i}{K\bar{r}} \right),
\end{equation}
where
\begin{equation} \label{eq:rbar_full_KM}
\bar{r} = \lim_{T\to \infty} \frac{1}{T}\int_0^T r(t) dt,
\end{equation}
is the time-averaged order parameter. For large $K$, (\ref{eq:mean_field_solution}) can be expanded to obtain $\phi_i \approx \frac{\omega_i}{K\bar{r}} + \mathcal{O}(K^{-3})$, which motivates the ansatz 
\begin{equation} \label{eq:linear_ansatz}
\hat{\phi}_i = \alpha \omega_i,
\end{equation}
used in previous studies \cite{Gottwald15, Gottwald17, HancockGottwald18, SmithGottwald19}. Note that $\alpha \sim 1/(K\bar{r})$. We refer to (\ref{eq:linear_ansatz}) as the \emph{linear} collective coordinate ansatz. Motivated by (\ref{eq:mean_field_solution}) we now consider the \emph{arcsin} collective coordinate ansatz
\begin{equation} \label{eq:asin_ansatz}
\hat{\phi}_i = \arcsin\left( \frac{ \omega_i}{K\alpha}\right).
\end{equation}
Here we use the order parameter $\bar{r}$ -- the only dynamical quantity in (4) -- as the shape parameter $\alpha$. For the arcsin ansatz (\ref{eq:asin_ansatz}), the evolution equation for the collective coordinate (\ref{eq:CC_general_form}) becomes
\begin{equation} \label{eq:asin_CC_evolution_equation}
 \dot{\alpha} = - \frac{K}{||\frac{d\hat{\bm{\phi}}}{d\alpha}||^2} \left(\sum_{i \in \mathcal{C}} \frac{s_i^2}{\sqrt{1-s_i^2}} \right) \left[ 1 - \frac{1}{N \alpha} \sum_{j\in \mathcal{C}} \sqrt{1-s_j^2} \right],
\end{equation}
where $s_i = s_i(\alpha,K) = \frac{\omega_i}{K\alpha}$. If $|\mathcal{C}| > 1$, there is a non-trivial cluster of synchronized oscillators, and the sum in round brackets in (\ref{eq:asin_CC_evolution_equation}) is positive. 
We remark that (\ref{eq:CC_general_form}) is true in general, whereas (\ref{eq:asin_CC_evolution_equation}) assumes that $\sum_{i\in \mathcal{C}} \omega_i = 0$. This is true for finite networks if the natural frequencies are symmetric about zero, as is the case for symmetric frequency distributions $g(\omega)$ with equiprobable sampling. However, it is generally not true that $\sum_{i\in \mathcal{C}} \omega_i = 0$ for random sampling from symmetric or non-symmetric frequency distributions. In such cases, (\ref{eq:asin_CC_evolution_equation}) can still be obtained by moving into the reference frame that rotates with the synchronized cluster $\mathcal{C}$, i.e., rotating with frequency $\Omega_\mathcal{C} = \sum_{i\in \mathcal{C}} \omega_i$, so that in this reference frame $\sum_{i\in \mathcal{C}} \omega_i = 0$. In this new reference frame the total mean frequency may not be zero.

Following from (\ref{eq:asin_CC_evolution_equation}), stationary points $\alpha^\star$ of the evolution equation (\ref{eq:asin_CC_evolution_equation}) correspond to solutions of
\begin{equation} \label{eq:asin_reduced_condition}
1 =   \frac{1}{N \alpha^\star} \sum_{j\in \mathcal{C}} \sqrt{1-s_j(\alpha^\star,K)^2}.
\end{equation}
This recovers the self-consistency equation for finite networks (cf. eq. (22) in \citet{RodriguesEtAl16} and eq. (3) in \citet{MirolloStrogatz05}). While self-consistency analysis yields the same stationary points as the collective coordinate approach (solutions to (\ref{eq:asin_reduced_condition})), the collective coordinate approach also yields dynamical information through the full evolution equation (\ref{eq:asin_CC_evolution_equation}). In particular, linear stability of the stationary points can be inferred easily from (\ref{eq:asin_CC_evolution_equation}). The dynamical nature of the collective coordinate approach also allows description of non-stationary attracting states, which occurs, for example, for multimodal natural frequency distributions\cite{Gottwald15, SmithGottwald19, MartensEtAl09, PazoErnest09, PietrasEtAl18}. From a computational standpoint, we note that (\ref{eq:asin_reduced_condition}) involves only a single sum, compared to the double sum in the general form of the collective coordinate approach (\ref{eq:CC_general_form}).

For both collective coordinate ansatzes, the solution $\hat{\bm{\phi}}(\alpha^\star)$ allows us to express the order parameter restricted to the ansatz manifold as
\begin{equation} \label{eq:rbar_CC}
 \bar{r}_\text{CC} = \frac{1}{N}\left| \sum_{j\in \mathcal{C}} e^{i \hat{\phi}_j(\alpha^\star)} \right|.
 \end{equation}
For the $\arcsin$ collective coordinate ansatz (\ref{eq:asin_ansatz}), the collective coordinate $\alpha$ replaces $\bar{r}$ in the mean field solution (\ref{eq:mean_field_solution}), and also satisfies the finite network self-consistency equation (\ref{eq:asin_reduced_condition}). Therefore, the arcsin ansatz is self-consistent in the sense that $\alpha^\star = \bar{r}_\text{CC}$, which is not true for the linear collective coordinate ansatz used in previous work \cite{Gottwald15, Gottwald17, HancockGottwald18, SmithGottwald19}.

We will show in the following section that both collective coordinate ansatzes accurately capture the collective dynamics and finite size effects of finite networks, with the $\arcsin$ ansatz yielding a significantly better approximation compared to the original collective coordinate ansatz (\ref{eq:linear_ansatz}).

\section{Performance of the collective coordinate framework for finite networks} \label{sec:CC-vs-OA}

We quantify the accuracy of the respective collective coordinate approaches by analyzing the differences between the order parameter $\bar{r}$ obtained from the full Kuramoto model (\ref{eq:rbar_full_KM}) and $\bar{r}_\text{CC}$ obtained from the respective collective coordinate reductions (\ref{eq:rbar_CC}) for several natural frequency distributions, as well as accuracy in identifying the synchronized cluster $\mathcal{C}$. 

We consider here two network sizes for the numerical simulation and the corresponding collective coordinate reduction of the Kuramoto model; a small network with $N=50$ where the finite size effects are clearly visible and a larger network with $N=500$ which qualitatively resembles the thermodynamic limit but still exhibits significant differences from it.

For all computations of the order parameter of the full Kuramoto model (\ref{eq:rbar_full_KM}) using direct numerical simulation, we use an adaptive fourth order Runge-Kutta scheme (\texttt{ode45} in MATLAB) with a maximum step size of $0.2$. We discard a transient of $7000$ time units, and use $2000$ time units for the time-averaging to ensure convergence.

\subsection{Order parameter}

To highlight the significance of finite size effects, and the ability of the collective coordinate methods to accurately capture them, we also show the order parameter obtained in the thermodynamic limit of infinitely many oscillators. By classical self-consistency analysis\cite{Strogatz00, AcebronEtAl05}, the order parameter in the thermodynamic limit satisfies for even frequency distributions $g(\omega)$
\begin{equation} \label{eq:therm_lim_self-consistency}
K \int_{-1}^{1} g(Kru) \sqrt{1-u^2} du = 1,
\end{equation}
which is the limit of (\ref{eq:asin_reduced_condition}) as $N\to \infty$, and can also be obtained via the Ott-Antonsen ansatz \cite{OmelchenkoWolfrum12, OmelchenkoWolfrum13}. We will show in Section~\ref{sec:therm_lim} that the self-consistency relationship (\ref{eq:therm_lim_self-consistency}) can also be derived from the arcsin collective coordinate ansatz in the thermodynamic limit (which follows from (\ref{eq:asin_reduced_condition})).

\subsubsection{Equiprobable draw of natural frequencies from a Lorentzian distribution}

\begin{figure}[tbp]
\centering
\includegraphics[width=\columnwidth]{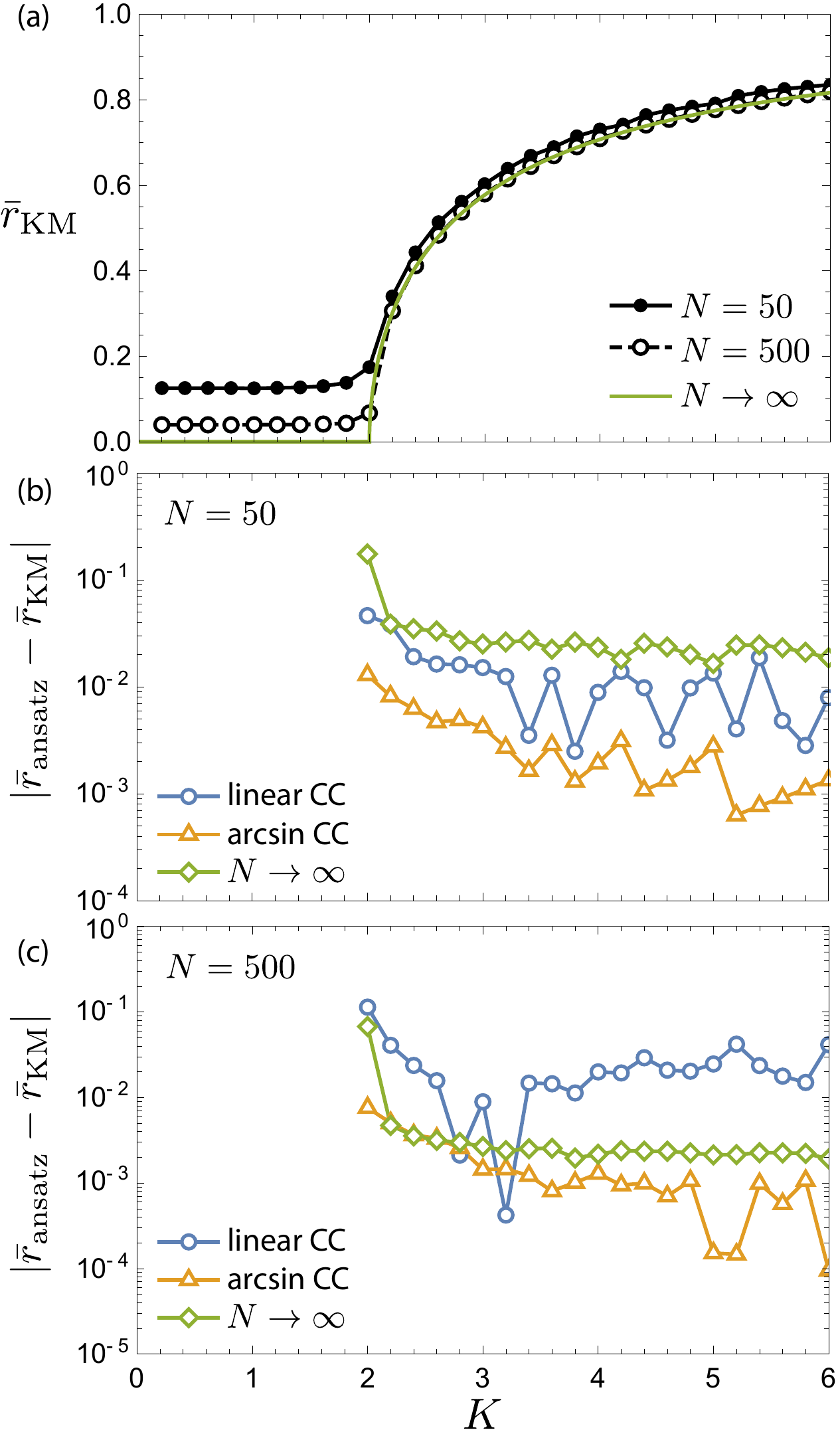}
\caption{(a)~Time-averaged order parameter $\bar{r}_\text{KM}$ for the Kuramoto model (\ref{eq:full_KM}) with $N=50$ (closed circles) and $N=500$ (open circles) oscillators with equiprobably drawn Lorentzian distributed natural frequencies (\ref{eq:lorentzian_distro}) with $\Delta=1$. The order parameter $r_\infty$ in the limit $N\to \infty$ (given by (\ref{eq:rbar_OA})) is shown in green. (b,c)~Error in the approximation $\bar{r}_\text{ansatz}$ obtained from the collective coordinate approaches (\ref{eq:rbar_CC}) compared to the full Kuramoto model $\bar{r}_\text{KM}$ [(b) $N=50$, (c) $N=500$]. Results are shown for the linear ansatz (\ref{eq:linear_ansatz}) (blue circles) and the $\arcsin$ ansatz (\ref{eq:asin_ansatz}) (orange triangles). The difference $|\bar{r}_\text{KM} - r_\infty|$ is shown by green diamonds to highlight finite size effects. The errors are shown for $K>K_c \approx 2$, when a synchronized cluster exists.}
\label{fig:Cauchy_rbar}
\end{figure}

We first consider a Lorentzian natural frequency distribution
\begin{equation} \label{eq:lorentzian_distro}
g(\omega) = \frac{\Delta}{\pi(\omega^2 + \Delta^2)},
\end{equation}
centered at zero with spread $\Delta>0$. In all simulations we choose $\Delta=1$. For the distribution (\ref{eq:lorentzian_distro}), the thermodynamic limit self-consistency equation (\ref{eq:therm_lim_self-consistency}) can be solved explicitly, yielding
 \begin{equation} \label{eq:rbar_OA}
 r_\infty =  \sqrt{1- \frac{2\Delta}{K}} \quad \text{if }K\geq 2 \Delta.
 \end{equation}
The solution (\ref{eq:rbar_OA}) of the self-consistency equation (\ref{eq:therm_lim_self-consistency}) is also found as the stationary solution of the evolution equation
\begin{equation} \label{eq:evolution_equation_OA}
\dot{r} =   - \Delta r + \frac{K}{2}\left( r - r^3 \right),
\end{equation}
derived via the Ott-Antonsen ansatz \cite{OttAntonsen08, OttAntonsen09}. The evolution equation (\ref{eq:evolution_equation_OA}) allows characterization of the stability of the order parameter. In particular, at $K_c = 2\Delta$ the incoherent state ($r=0$) loses stability and a partially synchronized state with $r=r_\infty$ emerges in a pitchfork bifurcation. This is shown by the green curve in Fig.~\ref{fig:Cauchy_rbar}(a).

For finite size networks, equiprobably drawn natural frequencies minimize finite size effects and are best suited to mimic the thermodynamic limit for finite but large $N$. Equiprobable draws are performed as follows. Let $G(\omega)$ denote the cumulative distribution function of the frequencies, then natural frequencies are drawn such that $G(\omega_j) = \frac{j}{N+1}$, for $j=1,\dots,N$. Fig.~\ref{fig:Cauchy_rbar}(a) shows $\bar{r}_\text{KM}$, estimated using (\ref{eq:rbar_full_KM}), as a function of the coupling strength $K$ for the full Kuramoto model (\ref{eq:full_KM}) with a small ($N=50$, closed circles) and a larger ($N=500$, open circles) number of oscillators. At $K_c \approx 2\Delta =  2$ there is a second order transition from the incoherent state, with $\bar{r}_\text{KM} \sim \mathcal{O}(1/\sqrt{N})$, to a partially synchronized state. The order parameter curves for $N=50$ and $N=500$ are very similar, albeit the curve for $N=500$ is smoother as it more closely represents the thermodynamic limit $r_\infty$ (\ref{eq:rbar_OA}), shown as the green curve. The effect of a finite size network, quantified as the difference between the finite network order parameter and $r_\infty$, is shown in Figs.~\ref{fig:Cauchy_rbar}(b,c) by the green diamonds. We see that for $N=50$ the differences are $\mathcal{O}(10^{-2})$ and for $N=500$ the differences are $\mathcal{O}(10^{-3})$. 

The error in the approximations $\bar{r}_\text{CC}$ compared to $\bar{r}_\text{KM}$ obtained from the full Kuramoto model (\ref{eq:full_KM}) with $N=50$ and $N=500$ are shown in Fig.~\ref{fig:Cauchy_rbar}(b) and Fig.~\ref{fig:Cauchy_rbar}(c), respectively, for the linear collective coordinate ansatz (\ref{eq:linear_ansatz}) and the $\arcsin$ collective coordinate ansatz (\ref{eq:asin_ansatz}). The errors are shown for $K>K_c \approx 2$, for which a synchronized cluster of oscillators exists. For $N=50$ (Fig.~\ref{fig:Cauchy_rbar}(b)), the $\arcsin$ ansatz (\ref{eq:asin_ansatz}) gives the best approximation for $\bar{r}$ (lowest error), and both collective coordinate ansatzes yield a more accurate approximation than the thermodynamic limit, and, hence, can be considered effective. For $N=500$ (Fig.~\ref{fig:Cauchy_rbar}(c)), the arcsin ansatz again yields the best approximation (with errors in the range $10^{-2} - 10^{-4}$), and less than half the error for $K\geq 3$ compared to the thermodynamic limit approximation. This indicates that a network of $N=500$ oscillators is not yet sufficiently large to be described by the thermodynamic limit. However, the thermodynamic limit yields a better approximation than the linear collective coordinate ansatz for $N=500$. We note that the pronounced dip in the error associated with the linear collective coordinate ansatz at $K\approx 3$ in Fig.~\ref{fig:Cauchy_rbar}(c) corresponds to a change in sign of $\bar{r}_\text{ansatz} - \bar{r}_\text{KM}$, i.e., the linear ansatz shifts from over-prediction to under-prediction of $\bar{r}$.

\subsubsection{Random draw of natural frequencies from a Lorentzian distribution}

 \begin{figure}[tbp]
\centering
\includegraphics[width=\columnwidth]{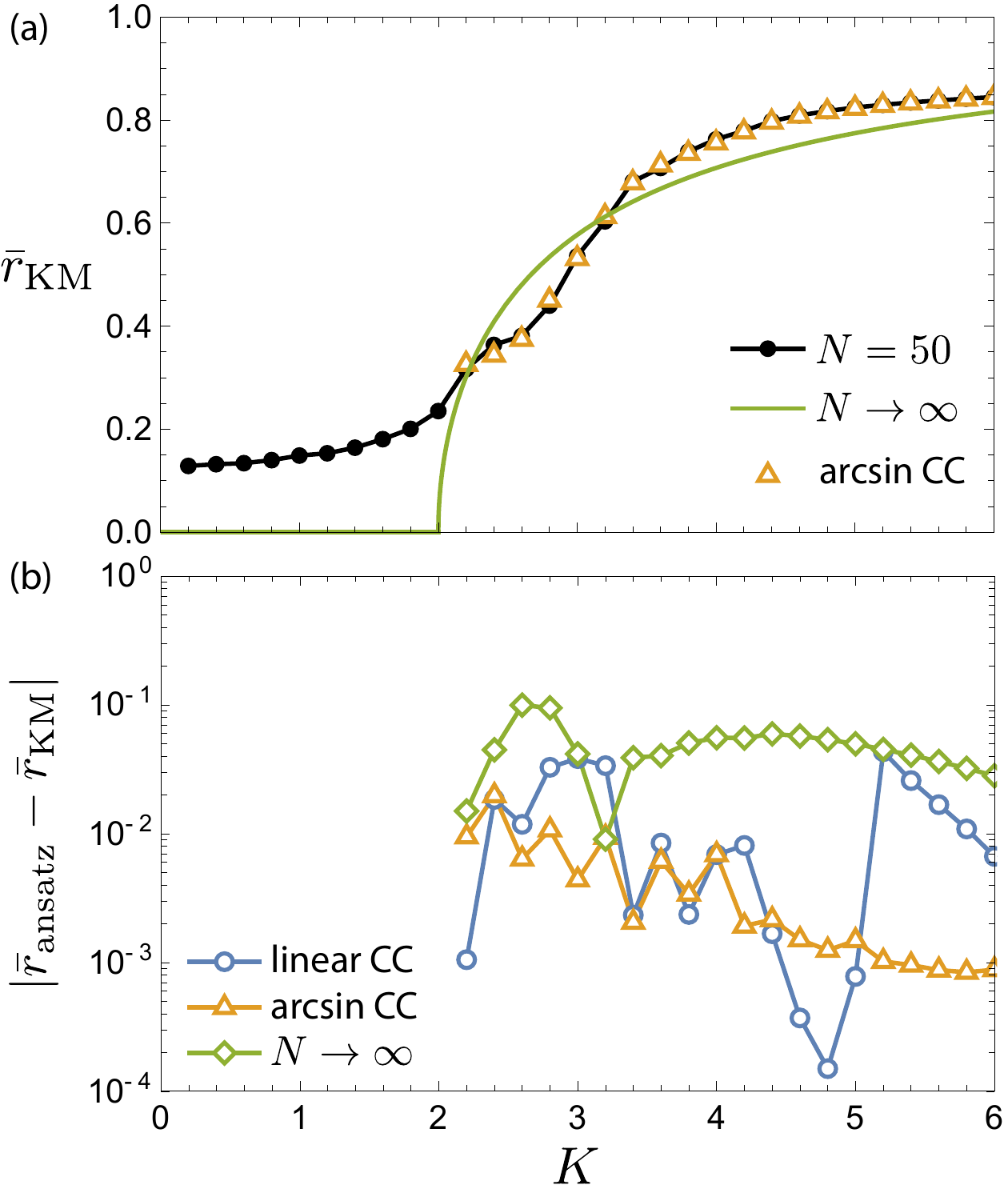}
\caption{(a)~Time averaged order parameter $\bar{r}_\text{KM}$ for the Kuramoto model (\ref{eq:full_KM}) with a single realization of $N=50$ oscillators with randomly drawn Lorentzian distributed natural frequencies (\ref{eq:lorentzian_distro}) with $\Delta=1$. The order parameter $r_\infty$ in the limit $N\to \infty$ (given by (\ref{eq:rbar_OA})) is shown in green and the collective coordinate approximations using the arcsin ansatz (\ref{eq:asin_ansatz}) are shown by the orange triangles.  (b)~Error in the approximation $\bar{r}_\text{ansatz}$ obtained from the collective coordinate approaches (\ref{eq:rbar_CC}) compared to the full Kuramoto model $\bar{r}_\text{KM}$. Results are shown for the linear ansatz (\ref{eq:linear_ansatz}) (blue circles) and the $\arcsin$ ansatz (\ref{eq:asin_ansatz}) (orange triangles). The difference $|\bar{r}_\text{KM} - r_\infty|$ is shown by green diamonds to highlight finite size effects. The errors are shown for $K\geq 2.2$, when a synchronized cluster with at least 10 oscillators exists.
}
\label{fig:random_Cauchy_rbar}
\end{figure}

We now show that the collective coordinate method, in particular the arcsin ansatz, accurately captures the collective dynamics when the natural frequencies are drawn randomly and finite size effects become exacerbated. Fig.~\ref{fig:random_Cauchy_rbar}(a) shows $\bar{r}_\text{KM}$ for the full Kuramoto model (\ref{eq:full_KM}) with $N=50$ oscillators with randomly drawn Lorentzian distributed frequencies for a single realization. Compared to equiprobably drawn frequencies (Fig.~\ref{fig:Cauchy_rbar}(a)), the  transition from the incoherent state to the partially synchronized state is not as well defined for randomly drawn frequencies (Fig.~\ref{fig:random_Cauchy_rbar}(a)). This is due to the existence of small synchronized clusters which gradually merge as $K$ increases. As one would expect, the order parameter in the thermodynamic limit (\ref{eq:rbar_OA})  (green curve in Fig.~\ref{fig:random_Cauchy_rbar}(a)) fails to capture finite size effects, such as non-monotonicity of the second derivative of $r(K)$. The collective coordinate ansatzes (\ref{eq:linear_ansatz}) and (\ref{eq:asin_ansatz}), on the other hand, reproduce the order parameter $\bar{r}_\text{KM}$ to high accuracy, again the arcsin ansatz being superior. The approximation given by the arcsin collective coordinate ansatz is shown by the orange diamonds in Fig.~\ref{fig:random_Cauchy_rbar}(a), where it is clearly seen that the approach captures finite size effects very well, such as non-monotonicity of the second derivative of $r(K)$ near the onset of partial synchronization. Analyzing in more detail, the error in reproducing the order parameter of the full Kuramoto model is shown in Fig.~\ref{fig:random_Cauchy_rbar}(b) for the collective coordinate ansatzes and the thermodynamic limit. The errors in $\bar{r}$ are shown for values $K\geq 2.2$, such that there exists a synchronized cluster with at least $10$ oscillators. For smaller values of $K$ there are several small synchronized clusters consisting of only a few oscillators, which interact. We ignore here this complicating issue but remark that the interaction of clusters can be approximated by a more complex collective coordinate ansatz \cite{Gottwald15, HancockGottwald18, SmithGottwald19}. We see clearly that the $\arcsin$ collective coordinate ansatz provides the best approximation for $\bar{r}$, and that both collective coordinate ansatzes generally yield a significantly more accurate approximation than the thermodynamic limit. 
As in Fig.~\ref{fig:Cauchy_rbar} the pronounced dips of the errors for both the thermodynamic limit and the linear collective coordinate ansatz are due to changes in the sign of $\bar{r}_\text{approx} - \bar{r}_\text{KM}$, i.e., changes from over-prediction to under-prediction, or vice versa, as can be clearly seen for the thermodynamic limit by the two intersections of the $\bar{r}_\text{KM}$ and $r_\infty$ curves in Fig.~\ref{fig:random_Cauchy_rbar}(a).

\subsubsection{Gaussian distributed natural frequencies}

\begin{figure}[tbp]
\centering
\includegraphics[width=\columnwidth]{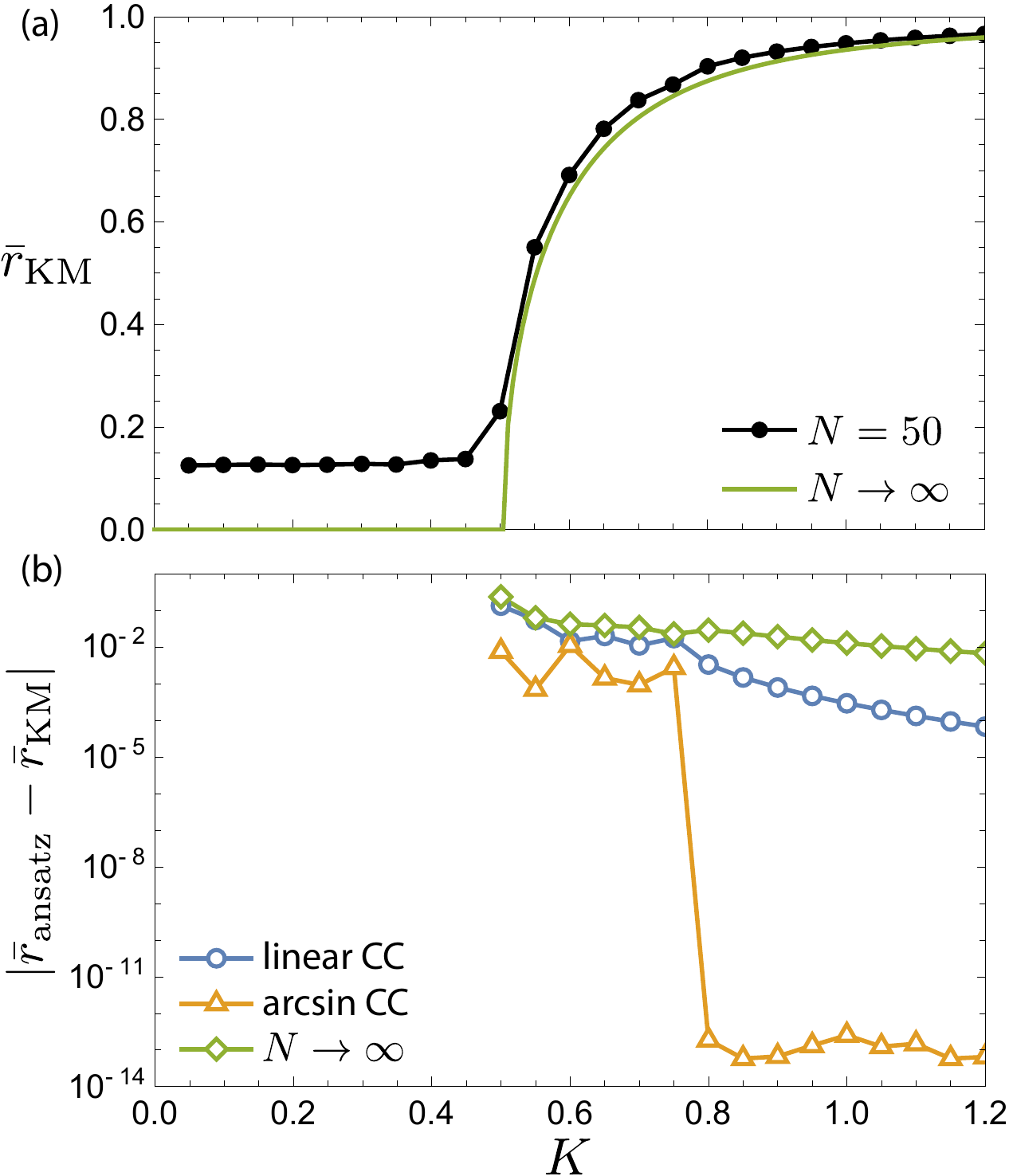}
\caption{(a)~Time averaged order parameter $\bar{r}_\text{KM}$ for the Kuramoto model (\ref{eq:full_KM}) with $N=50$ oscillators with equiprobable Gaussian distributed natural frequencies (mean zero, and variance $\sigma^2 = 0.1$). The order parameter $r_\infty$ in the limit $N\to \infty$ (given as solution of (\ref{eq:therm_lim_Gaussian})) is shown in green. (b)~Difference between $\bar{r}_\text{KM}$ obtained from the full Kuramoto model (\ref{eq:rbar_full_KM}) and $\bar{r}_\text{CC}$ obtained from collective coordinate ansatzes (\ref{eq:rbar_CC}). Results are shown for the linear ansatz (\ref{eq:linear_ansatz}) (blue circles) and the $\arcsin$ ansatz (\ref{eq:asin_ansatz}) (orange triangles). The difference $|\bar{r}_\text{KM} - r_\infty|$ is shown by green diamonds to highlight finite size effects. The differences are shown for $K\geq K_c \approx 0.5$, when a synchronized cluster exists.}
\label{fig:Gaussian_rbar}
\end{figure}

For Gaussian distributed natural frequencies with mean zero and variance $\sigma^2$ the self-consistency equation in the thermodynamic limit (\ref{eq:therm_lim_self-consistency}) becomes
\begin{equation} \label{eq:therm_lim_Gaussian}
\frac{\sqrt{\pi}\, K \exp\left(-\frac{K^2r^2}{4\sigma^2}\right)}{2\sqrt{2}\sigma} \left( I_0\left(\frac{K^2r^2}{4\sigma^2}\right) + I_1\left( \frac{K^2r^2}{4\sigma^2} \right) \right) = 1,
\end{equation}
where $I_n(z)$ denotes Bessel functions of the first kind. This implicit equation can be solved numerically to obtain $r_\infty(K)$ for any value of $K$, shown as the green curve in Fig.~\ref{fig:Gaussian_rbar}(a) for $\sigma^2 = 0.1$. The critical coupling strength $K_c$ corresponding to the onset of partial synchronization can be found by substituting $r=0$ into (\ref{eq:therm_lim_Gaussian}), yielding 
\begin{equation} \nonumber
K_c = 2\sigma \sqrt{2/\pi},
\end{equation}
which amounts for $K_c \approx 0.5$ for $\sigma^2 = 0.1$.

As for Lorentzian distributed natural frequencies, we compare the error in reproducing the order parameter $\bar{r}_\text{KM}$ associated with the linear and the arcsin collective coordinate ansatzes, as well as the difference $|\bar{r}_\text{KM} - r_\infty|$ to illustrate the significance of finite size effects. The results are shown in Fig.~\ref{fig:Gaussian_rbar}(b) for $N=50$ oscillators with equiprobably drawn natural frequencies. We see that both collective coordinate ansatzes accurately capture finite size effects, since they yield smaller errors than the thermodynamic limit. The arcsin ansatz yields the best approximation. In fact, for values of $K\geq K_g$, where $K_g$ denotes the onset of global synchronization when all $N$ oscillators synchronize ($K_g\approx 0.8$), the difference between $\bar{r}_\text{CC}$ obtained from the $\arcsin$ collective coordinate approach compared to $\bar{r}_\text{KM}$ obtained from the full Kuramoto model (\ref{eq:rbar_full_KM}) is $\mathcal{O}(10^{-13})$, corresponding to the numerical precision of the computational methods. This is because the arcsin ansatz is both exact for globally synchronized states and self-consistent in the sense that the static collective coordinate $r$ is equal to the order parameter $\bar{r}$.

\subsubsection{Uniformly distributed natural frequencies}

\begin{figure}[tbp]
\centering
\includegraphics[width=\columnwidth]{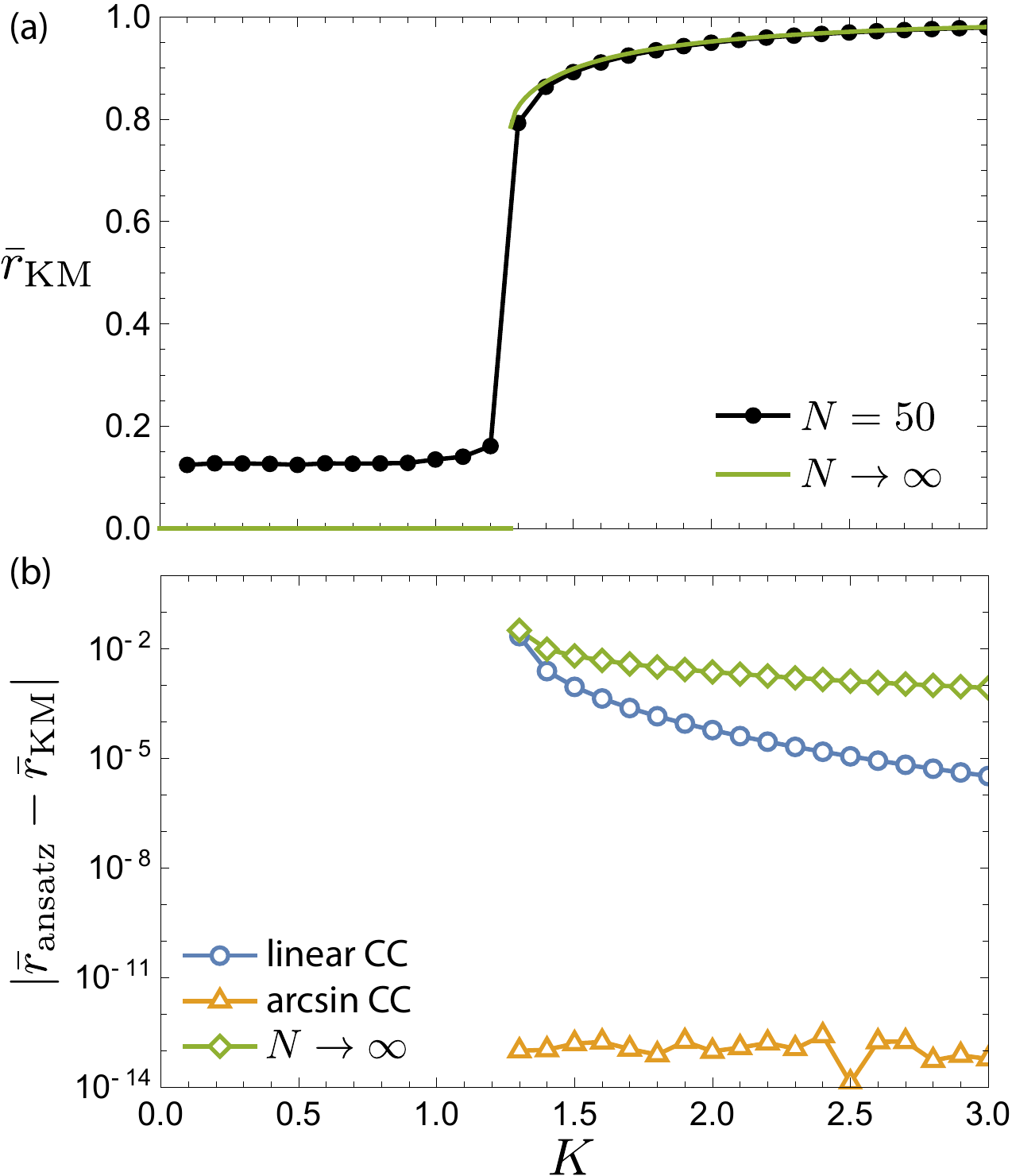}
\caption{(a)~Time averaged order parameter $\bar{r}_\text{KM}$ for the Kuramoto model (\ref{eq:full_KM}) with $N=50$ oscillators with equiprobable uniformly distributed natural frequencies between $-1$ and $1$. The order parameter $r_\infty$ in the limit $N\to \infty$ (given as solution of (\ref{eq:therm_lim_uniform})) is shown in green. (b)~Difference between $\bar{r}_\text{KM}$ obtained from the full Kuramoto model (\ref{eq:rbar_full_KM}) and $\bar{r}_\text{CC}$ obtained from collective coordinate ansatzes (\ref{eq:rbar_CC}). Results are shown for the linear ansatz (\ref{eq:linear_ansatz}) (blue circles) and the $\arcsin$ ansatz (\ref{eq:asin_ansatz}) (orange triangles). The difference $|\bar{r}_\text{KM} - r_\infty|$ is shown by green diamonds to highlight finite size effects. The differences are shown for $K \geq K_c \approx 1.3$, when a synchronized cluster exists.}
\label{fig:Uniform_rbar}
\end{figure}

For uniformly distributed natural frequencies on the interval $[-a,a]$ the self-consistency equation in the thermodynamic limit (\ref{eq:therm_lim_self-consistency}) becomes
\begin{equation} \label{eq:therm_lim_uniform} 
1  = \begin{cases}
\frac{\pi K}{4a} &\text{if } Kr \leq a \\
\frac{K}{2}\left(\frac{\sqrt{K^2 r^2 - a^2}}{K^2r^2} + \frac{1}{a} \text{arccsc}\left(\frac{Kr}{a}\right) \right) &\text{if } Kr > a
\end{cases}.
\end{equation}
This implies an explosive first-order transition from incoherence to global synchronization\cite{Pazo05}, with $K_c = 4a/\pi$, and the order parameter at the transition is $r_c = \pi/4$. This explosive transition in the thermodynamic limit is shown in Fig.~\ref{fig:Uniform_rbar}(a) by the green curve.

Fig.~\ref{fig:Uniform_rbar}(b) shows that, like the Gaussian and Lorentzian frequency distributions, both the linear and arcsin collective coordinate ansatzes yield good approximations for the collective dynamics and accurately capture finite size effects, with the arcsin ansatz performing significantly better. For the case of uniformly distributed natural frequencies, the arcsin ansatz is exact for all values of $K>K_c \approx 1.3$ because the transition to synchronization is explosive and all oscillators synchronize at $K=K_c$, i.e. $K_g = K_c$, leading to errors of $\mathcal{O}(10^{-13})$, corresponding to the numerical precision of the computational methods.

\subsection{The synchronized cluster $\mathcal{C}$} \label{sec:cluster}

Along with accurately predicting the order parameter, the collective coordinate approach has the advantage of being able to predict the set of oscillators $\mathcal{C}$ that will synchronize. The set of synchronized oscillators $\mathcal{C}$ is identified within the collective coordinate approach as the maximal set of oscillators such that (\ref{eq:CC_general_form}) has a stable stationary solution $\alpha^\star$. The set $\mathcal{C}$ can be found by starting at a high value of $K$ such that most oscillators synchronize, and then successively removing oscillators from $\mathcal{C}$ whenever stationary solutions of (\ref{eq:CC_general_form}) cease to exist (resulting from saddle-node bifurcations that will be discussed in Section~\ref{sec:SN_bifurcation}). In the case of all-to-all coupling, which is what we consider here, when decreasing $K$, the oscillator that becomes desynchronized is always the oscillator with natural frequency furthest from the mean frequency of the current synchronized cluster, $\Omega_\mathcal{C} = (1/|\mathcal{C}|)\sum_{j\in \mathcal{C}} \omega_j$. With this criterion, oscillators can also be added to the synchronized cluster when increasing $K$ by including the non-entrained oscillator with natural frequency closest to the mean frequency of the synchronized cluster. For more complex network topologies, the oscillators that become desynchronized can be identified within the collective coordinate framework by linearizing the full Kuramoto model (\ref{eq:full_KM}) around the ansatz solution $\hat{\bm{\phi}}(\alpha^\star)$ \cite{HancockGottwald18}.

\begin{figure}[tbp]
\centering
\includegraphics[width=\columnwidth]{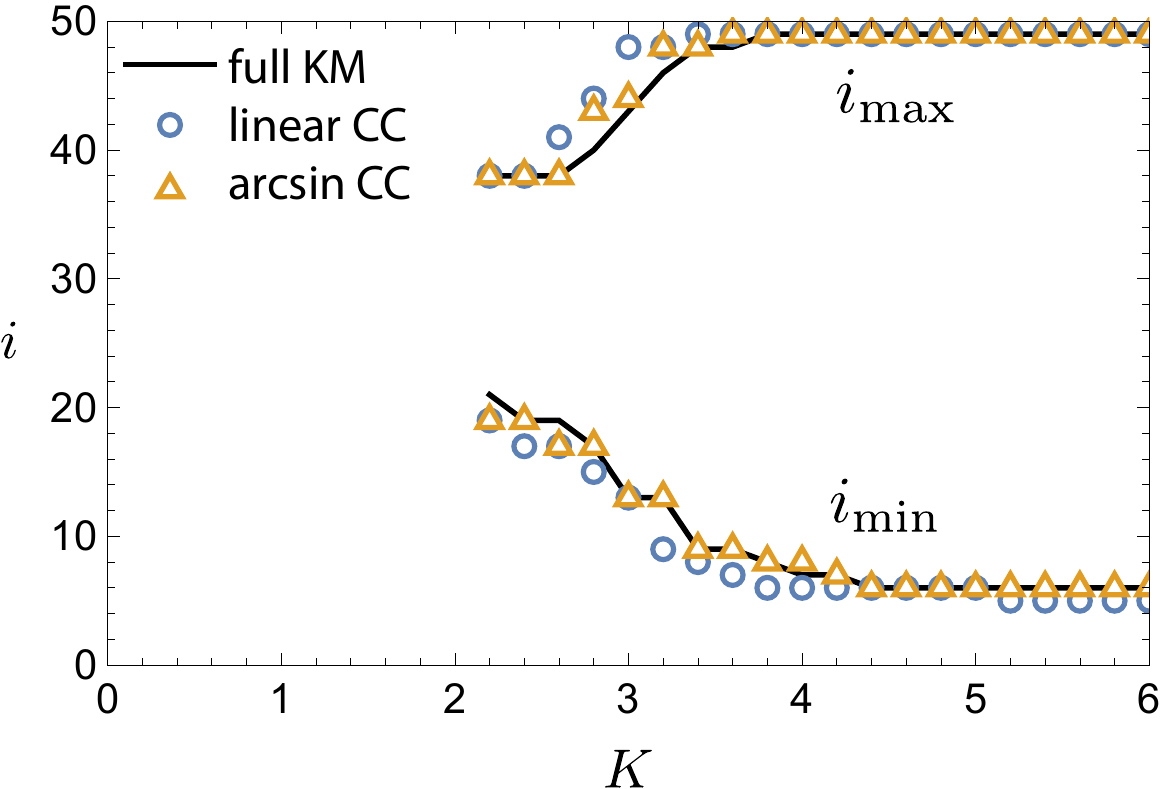}
\caption{Lower ($i_{\min}$) and upper ($i_{\max}$) boundaries of the synchronized cluster for $N=50$ oscillators with randomly drawn Lorentzian distributed natural frequencies (as in Fig.~\ref{fig:random_Cauchy_rbar}). Results are shown for the full Kuramoto model (solid black), and for collective coordinates [linear ansatz (\ref{eq:linear_ansatz}) (blue circles) and $\arcsin$ ansatz (\ref{eq:asin_ansatz}) (orange triangles)], for $K\geq 2.2$, when a synchronized cluster with at least 10 oscillators exists.}
\label{fig:Cauchy_C}
\end{figure}

 For the full Kuramoto model (\ref{eq:full_KM}) the set $\mathcal{C}$ is found by computing the effective frequency of each oscillator,
 \begin{equation} \nonumber
 \bar{\omega}_i = \lim_{T\to \infty} \frac{1}{T}\int_0^T \dot{\phi}_i dt = \lim_{T\to \infty} \frac{\phi_i(T) - \phi_i(0) + 2\pi w_i(T)}{T},
 \end{equation}
 where $w_i(T)\in \mathbb{Z}$ is the winding number of the $i$-th oscillator. Synchronized clusters are sets of oscillators with the same effective frequency, and $\mathcal{C}$ is the largest such synchronized cluster. Labeling the oscillators in order of increasing natural frequencies in an all-to-all coupled network, the synchronized cluster consists of all oscillators with indices between some minimum index $i_{\min}$ and some maximum index $i_{\max}$. Hence, the size of the synchronized cluster is $i_{\max} - i_{\min} + 1$. Fig.~\ref{fig:Cauchy_C} shows the lower ($i_{\min}$) and upper ($i_{\max}$) boundaries of the cluster for $N=50$ oscillators with randomly drawn Lorentzian distributed natural frequencies (the same as in Fig.~\ref{fig:random_Cauchy_rbar}). As expected, the synchronized cluster grows in size monotonically upon increasing the coupling strength $K$, with the lower boundary $i_{\min}$ decreasing monotonically and the upper boundary $i_{\max}$ increasing monotonically. We observe that the $\arcsin$ collective coordinate ansatz (\ref{eq:asin_ansatz}) (orange triangles) agrees with the full Kuramoto model (solid black curve) for most values of $K$, whereas the linear collective coordinate ansatz (\ref{eq:linear_ansatz}) generally overpredicts the size of the synchronized cluster.
 
\section{Cascade of saddle-node bifurcations} \label{sec:SN_bifurcation}

\begin{figure}[tbp]
\centering
\includegraphics[width=\columnwidth]{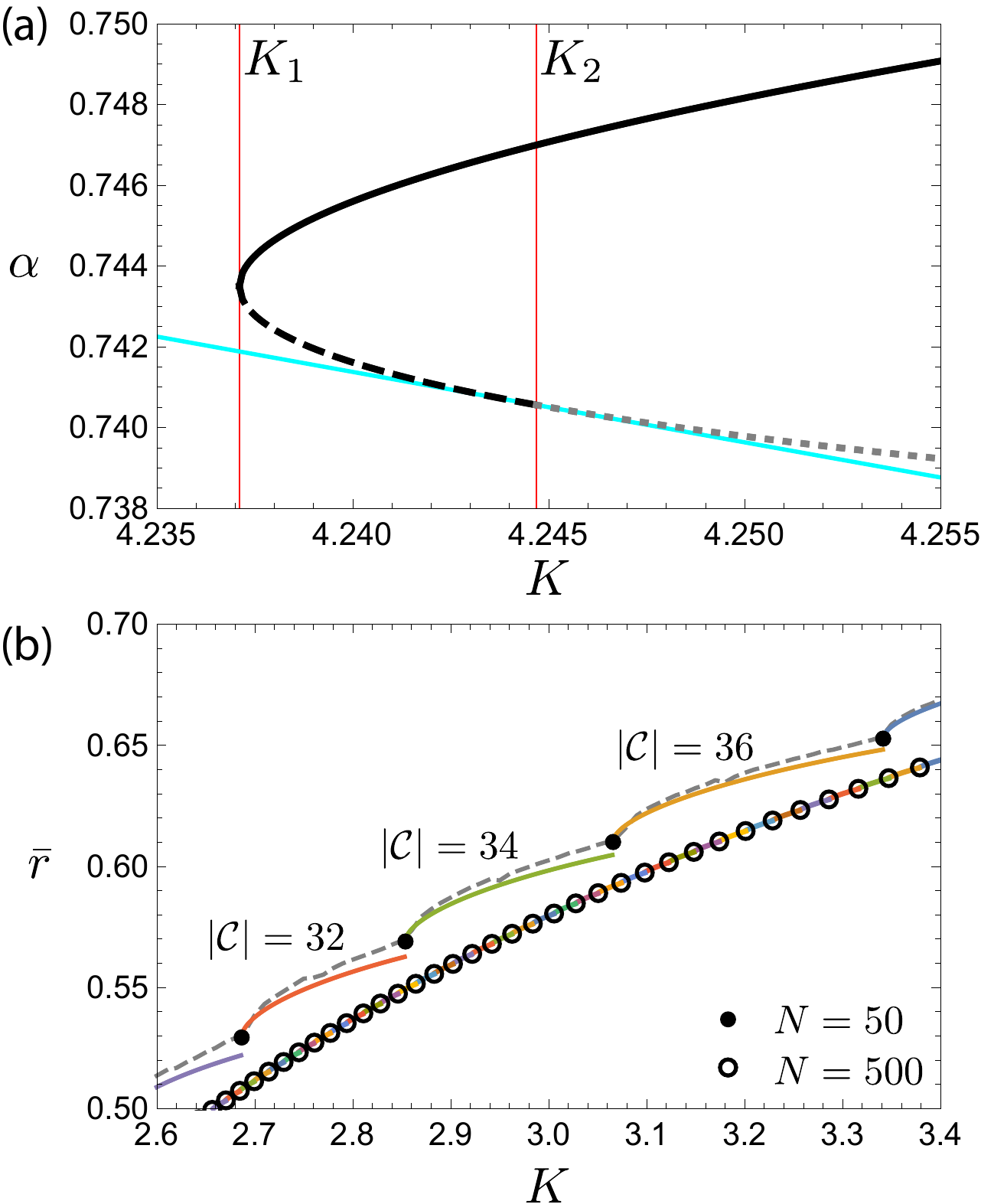}
\caption{(a)~Stable (solid black) and unstable (dashed black and dotted gray) stationary solutions of (\ref{eq:asin_CC_evolution_equation}) annihilate via a saddle-node bifurcation at $K=K_1 \approx 4.237$. For $K>K_2 \approx 4.245$ the unstable solution switches to the complex branch such that $\sqrt{1-s_{j^*}^2}$ is negative, where $j^*$ is such that $\omega_{j^*} = \max_{i\in \mathcal{C}} |\omega_i|  \eqqcolon \omega_\mathcal{C}$. At $K=K_2$ the unstable stationary solution intersects tangentially with the curve $\alpha = \omega_\mathcal{C}/K$ (cyan). (b)~Cascades of saddle-node bifurcations for $N=50$ and $N=500$. The bifurcation points are marked by closed circles and open circles, respectively. The stable stationary solutions of (\ref{eq:asin_CC_evolution_equation}) are shown as solid colored curves. The order parameter $\bar{r}_\text{KM}$ of the full Kuramoto model is shown by the dashed gray curve for $N=50$. Both (a) and (b) consider equiprobable Lorentzian distributed natural frequencies.}
\label{fig:Cauchy_saddle-node_1}
\end{figure}

For the linear ansatz (\ref{eq:linear_ansatz}), and a set of oscillators $\mathcal{C}$, stationary points of (\ref{eq:CC_general_form}), if they exist, form a pair (one stable and one unstable) which annihilate via a saddle-node bifurcation at $K=K_1(\mathcal{C})$ \cite{Gottwald15}. For $\mathcal{C}$ consisting of all oscillators, i.e., global synchronization, this agrees with the analysis of Mirollo and Strogatz \cite{MirolloStrogatz05} which showed that the transition from global synchronization to partial synchronization occurs as a saddle-node bifurcation for finite networks.  As for the linear collective coordinate ansatz, for each set of oscillators $\mathcal{C}$ a saddle-node bifurcation occurs for the arcsin ansatz (\ref{eq:asin_ansatz}), as shown in Fig.~\ref{fig:Cauchy_saddle-node_1}(a) for $N=50$ oscillators with equiprobable Lorentzian distributed natural frequencies, with $\mathcal{C}$ consisting of the $|\mathcal{C}|=42$ oscillators with natural frequencies closest to the (zero) mean frequency. For some range of coupling strengths $K>K_1$, $\mathcal{C}$ forms the synchronized cluster. We note that for the arcsin ansatz, at $K=K_2(\mathcal{C})>K_1$ the unstable stationary point of (\ref{eq:asin_CC_evolution_equation}) satisfies 
\begin{equation*}
s_{j^*} (\alpha^\star,K_2) = \frac{\omega_{j^*}}{K_2\,\alpha^\star}= 1,
\end{equation*}
where $j^*$ is such that $\omega_{j^*} = \max_{i\in \mathcal{C}} |\omega_i|  \eqqcolon \omega_\mathcal{C}$. This means that at $K=K_2$ the unstable solution curve (dashed black) intersects tangentially with the curve $\alpha = \omega_\mathcal{C}/K$ (cyan curve in Fig.~\ref{fig:Cauchy_saddle-node_1}). For $\alpha < \omega_\mathcal{C}/K$ (below the cyan curve) the evolution equation for the collective coordinate (\ref{eq:asin_CC_evolution_equation}) is complex valued, and has no physical meaning.  For $K>K_2$, the unstable solution of (\ref{eq:asin_CC_evolution_equation}) switches to a different complex branch (dotted gray) such that $\sqrt{1-s_{j^*}^2}$ is negative in (\ref{eq:asin_CC_evolution_equation}). 

For both collective coordinate ansatzes, stable stationary points of (\ref{eq:CC_general_form}) may be found for $K<K_1(\mathcal{C})$ for smaller subsets $\mathcal{C}'\subset \mathcal{C}$, and the bifurcation sequence repeats. Therefore, for both collective coordinate ansatzes, the transition from global synchronization to partial synchronization, and then to incoherence, as $K$ decreases, occurs for unimodal frequency distributions as a cascade of saddle-node bifurcations, successively removing more and more oscillators from the synchronized set $\mathcal{C}$. This cascade is shown for the arcsin collective coordinate ansatz in Fig.~\ref{fig:Cauchy_saddle-node_1}(b), where the stable stationary solution branches of (\ref{eq:asin_CC_evolution_equation}) are shown as solid curves for a range of cluster sizes with the saddle-node bifurcations marked by black points at the left end of each solution branch. We terminate solution branches for smaller cluster sizes at the saddle-node bifurcation of larger cluster sizes, e.g., the solution branch with $|\mathcal{C}| = 34$ terminates at $K\approx 3.07$ where the branch with $|\mathcal{C}|=36$ begins. This follows from the assumption that all oscillators that can synchronize will synchronize, and so the steady state of the system will have maximal synchronized cluster size $|\mathcal{C}|$. Due to symmetry of the natural frequencies, oscillators are successively removed from $\mathcal{C}$ in pairs as the coupling strength decreases. We note that the collective coordinate approach reproduces the order parameter of the full Kuramoto model (dashed gray in Fig.~\ref{fig:Cauchy_saddle-node_1}(b)) most accurately at each saddle-node bifurcation, with gradual deviation from the full Kuramoto model as $K$ increases away from each saddle-node bifurcation. This phenomenon is also evident in Fig.~\ref{fig:Cauchy_rbar}(b), where for $K>3.5$ the error of the arcsin collective coordinate ansatz experiences sharp decreases followed by gradual increases. The collective coordinate approach clearly captures finite size effects such as non-monotonicity of the second derivative of $r(K)$, with points of inflection at each saddle-node bifurcation. This cascade of saddle-node bifurcations is consistent with the onset of synchronization being a second order phase transition for unimodal frequency distributions. We note that the saddle-node bifurcation is only for the order parameter. The actual Kuramoto model (\ref{eq:full_KM}) undergoes a complex bifurcation from one (possibly chaotic) attractor to another one.

A similar cascade has been studied by Paz\'{o} \cite{Pazo05} for frequency distributions with compact supports (such as uniform distributions), where there is a cascade of frequency splittings, such that the globally synchronized cluster splits into multiple smaller clusters, each having a different effective frequency. This frequency splitting cascade could also be described by the collective coordinate approach, albeit with a more complex ansatz function that allows for multiple synchronized clusters \cite{Gottwald15, HancockGottwald18, SmithGottwald19}.

\section{Collective coordinate reduction in the thermodynamic limit} \label{sec:therm_lim}

We now show analytically that in the thermodynamic limit the $\arcsin$ collective coordinate ansatz reproduces the same results for the bifurcation structure of the order parameter as the Ott-Antonsen ansatz (\ref{eq:rbar_OA}), and recovers well-known relations between the coupling strength and the order parameter for partially synchronized states for general natural frequency distributions. In addition, the collective coordinate framework provides dynamical information for the evolution of small perturbations of the synchronized state along the ansatz manifold, which is not captured by the Ott-Antonsen ansatz (\ref{eq:evolution_equation_OA}).

 Taking the limit as $N\to \infty$ in the evolution equation for the collective coordinate (\ref{eq:CC_general_form}) involves replacing summations with integrals, i.e.,
\begin{equation} \nonumber
\lim_{N\to \infty} \frac{1}{N} \sum_{i\in \mathcal{C}} h(\omega_i) \longrightarrow \int_{-\omega_\mathcal{C}(\infty)}^{\omega_\mathcal{C}(\infty)} h(\omega) g(\omega) d\omega,
\end{equation}
for some function $h$, where $\omega_\mathcal{C}(\infty) = \lim_{N\to \infty} \omega_\mathcal{C}(N)$ and $\omega_\mathcal{C}(N) =  \max_{i\in \mathcal{C}} \left|\omega_i\right|$. For the $\arcsin$ ansatz (\ref{eq:asin_ansatz}), $\omega_\mathcal{C}(\infty) = K\alpha$ (since the domain of $\arcsin$ is $[-1,1]$). For finite networks, there are finitely many coupling strengths $K_1(\mathcal{C})$ that correspond to saddle-node bifurcations of the collective coordinate dynamics (\ref{eq:asin_CC_evolution_equation}) (cf. Fig.~\ref{fig:Cauchy_saddle-node_1}) because each saddle-node bifurcation corresponds to removing oscillators from $\mathcal{C}$, and there are only finitely many oscillators that can be removed. As $N$ increases, the intervals between successive saddle-node bifurcations decreases, as demonstrated in Fig.~\ref{fig:Cauchy_saddle-node_1}(b) by comparing $N=50$ to $N=500$. In the thermodynamic limit the intervals between saddle-node bifurcations converge to zero, so that saddle-node bifurcations occur at every coupling strength $K$, with oscillators satisfying $|\omega| = \omega_\mathcal{C}$ being neutrally stable in the synchronized set $\mathcal{C}$.

In the thermodynamic limit the evolution equation for the collective coordinate (\ref{eq:CC_general_form}) becomes
\begin{equation} \label{eq:CC_thermo_general_form}
I_3(\alpha,K) \,\dot{\alpha} = I_1(\alpha,K) + K I_2(\alpha,K),
\end{equation}
where 
\begin{align}
I_1(\alpha,K) &= \lim_{N\to \infty} \frac{1}{N}\left\langle \bm{\omega}, \frac{d\hat{\bm{\phi}}}{d\alpha}\right\rangle \nonumber
\\
I_2(\alpha,K) &= \lim_{N\to \infty} \frac{1}{N^2}\sum_{i,j\in \mathcal{C}} \frac{d\hat{\phi_i}}{d\alpha} \sin(\hat{\phi}_j - \hat{\phi}_i)  \nonumber
\\
I_3(\alpha,K) &= \lim_{N\to \infty} \frac{1}{N} \left\langle \frac{d\hat{\bm{\phi}}}{d\alpha} , \frac{d\hat{\bm{\phi}}}{d\alpha}\right\rangle .  \nonumber
\end{align}
Recall that for the arcsin ansatz (\ref{eq:asin_ansatz}) we have $\alpha(t) = r(t)$. In this case (\ref{eq:CC_thermo_general_form}) can be expanded to yield
\begin{align} 
I_3 \dot{r} = \int_{-Kr}^{Kr} & g(\omega) \frac{d\hat{\phi}}{dr} \omega \, \times \nonumber  \\
& \left[ 1 + \frac{K}{\omega} \int_{-Kr}^{Kr} g(\eta) \sin\left(\hat{\phi}(\eta) - \hat{\phi}(\omega) \right) d\eta \right] d\omega.   \label{eq:therm_lim_1}
\end{align}
Making the change of variables $s = \frac{\omega}{Kr}$ and $u=\frac{\eta}{Kr}$, for the $\arcsin$ collective coordinate ansatz (\ref{eq:asin_ansatz}), equation (\ref{eq:therm_lim_1}) becomes
\begin{equation} \label{eq:therm_lim_2}
I_3 \dot{r}= - J_1(r,K)\, r \left(1 - J_2(r,K) \right),
\end{equation}
where 
\begin{align}
J_1(r,K) &= K^2 \int_{-1}^{1} \frac{g(Krs)  s^2}{\sqrt{1-s^2}} ds, \label{eq:therm_lim_int_1}   \\ 
J_2(r,K) &= K \int_{-1}^{1} g(Kru) \sqrt{1-u^2} du.  \label{eq:therm_lim_int_2}
\end{align}
Note that $J_1>0$ provided that the synchronized set of oscillators (i.e., those satisfying $|\omega|<Kr$) has non-zero measure. Stationary solutions $r^\star$ of (\ref{eq:therm_lim_2}) are given by $r^\star = 0$ and as the solutions of 
\begin{equation} \label{eq:therm_lim_condition}
J_2(r^\star,K) = 1.
\end{equation}
This recovers the well-known self-consistency equation (\ref{eq:therm_lim_self-consistency}) for the Kuramoto model in the thermodynamic limit\cite{Strogatz00, AcebronEtAl05, OmelchenkoWolfrum12, OmelchenkoWolfrum13}. As such, through the collective coordinate approach we recover the relations $r(K)$ described explicitly by (\ref{eq:rbar_OA}) for Lorentzian distributed natural frequencies, and described implicitly by (\ref{eq:therm_lim_Gaussian}) and (\ref{eq:therm_lim_uniform}) for Gaussian and uniformly distributed natural frequencies, respectively.

We remark that while the Ott-Antonsen ansatz \cite{OttAntonsen08, OttAntonsen09} also yields an evolution equation for the order parameter in the thermodynamic limit, the Ott-Antonsen approach is only applicable to analytic frequency distributions, whereas the collective coordinate evolution equation (\ref{eq:CC_thermo_general_form}) can be applied to any natural frequency distribution. For uniform distributions, which are not analytic, the dynamics can be approximated by the Ott-Antonsen approach by approximating the distribution by a sequence of rational functions \cite{Skardal18}.

We now study the evolution equation for the collective coordinate in the thermodynamic limit for Lorentzian distributed natural frequencies. We shall see that besides recovering the results from the Ott-Antonsen ansatz (\ref{eq:rbar_OA}) and mean field theory (\ref{eq:therm_lim_condition}) on the stationary points, the collective coordinate approach also encapsulates dynamical information on the behavior of small perturbations.

\subsection{Lorentzian natural frequency distribution}

For a Lorentzian distribution (\ref{eq:lorentzian_distro}), the evolution equation (\ref{eq:therm_lim_2}) reduces to 
\begin{equation} \label{eq:CC_thermo_Lorentzian}
\epsilon^{-1} \dot{r} =   - \Delta r + \frac{K}{2}\left( r - r^3 \right),
\end{equation}
where $\epsilon = D/I_3$ with
\begin{equation}
D =\frac{ 2\Delta}{E\left(\Delta + E \right) \left( Kr^2 + \Delta + E\right)} \text{ and }
E = \sqrt{K^2r^2 + \Delta^2}. \nonumber
\end{equation}

\subsubsection{Bifurcation structure}

The right hand side of (\ref{eq:CC_thermo_Lorentzian}) is identical to that obtained via the Ott-Antonsen approach \cite{OttAntonsen08, OttAntonsen09}. Therefore, the collective coordinate approach recovers the same pitchfork bifurcation occurring at $K = 2\Delta$, such that for $K< 2\Delta$ the incoherent state $r=0$ is stable, and for $K>2\Delta$ the incoherent state is unstable, and a stable synchronized state emerges, with $r$ given by (\ref{eq:rbar_OA}) (cf. Fig.~\ref{fig:Cauchy_rbar}(a) and Fig.~\ref{fig:random_Cauchy_rbar}(a)).

\subsubsection{Relaxation rate toward the synchronized state}

\begin{figure}[tbp]
\centering
\includegraphics[width=\columnwidth]{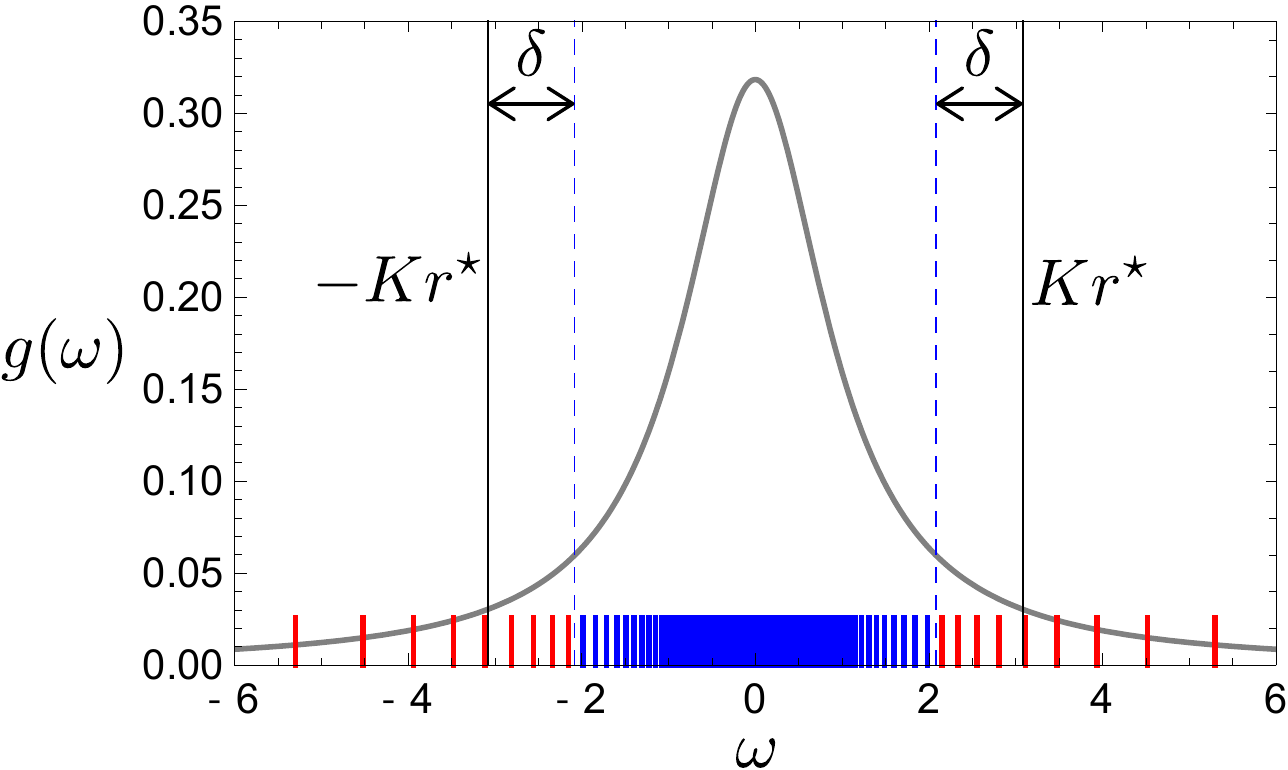}
\caption{Lorentzian natural frequency distribution with $\Delta=1$. The limits $\pm\omega_{\mathcal{C}}=\pm Kr^\star$ are shown for $K=4.24$, with $r^\star$ given by (\ref{eq:rbar_OA}). $N=100$ equiprobably drawn natural frequencies are shown as ticks on the horizontal axis, colored blue for $\omega_i \in [-Kr^\star+\delta,Kr^\star-\delta]$ with $\delta = 1$, and red otherwise.}
\label{fig:rel_rate_cluster}
\end{figure}

\begin{figure}[tbp]
\centering
\includegraphics[width=\columnwidth]{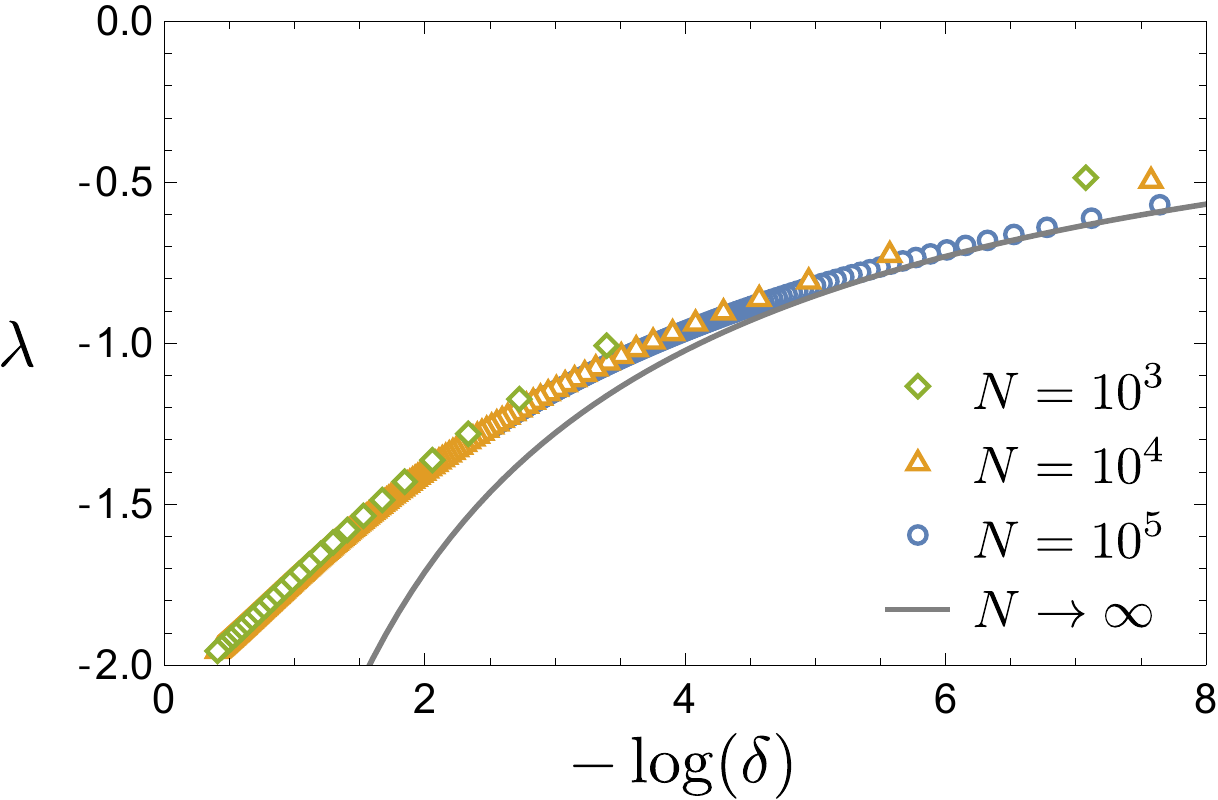}
\caption{Relaxation rate $\lambda$ toward the stationary solution $r^\star$ along the ansatz manifold $\hat{\bm{\phi}}(r)$ for $K=4.24$, given by (\ref{eq:CC_therm_lim_rel_rate_delta}) for $N\to \infty$ (solid black curve) and given by (\ref{eq:finite_CC_relaxation_rate}) for finite $N$, where the cluster $\mathcal{C} = \mathcal{C}'(\delta,N)$ is determined by (\ref{eq:finite_CC_delta_cluster}).}
\label{fig:rel_rate_vs_delta}
\end{figure}

In the limit as $\omega_\mathcal{C} \to Kr$, the integral $I_3 \to \infty$, and, hence, $\epsilon \to 0$. This has a physical interpretation linked to the cascade of saddle-node bifurcations discussed in Section~\ref{sec:SN_bifurcation} and illustrated in Fig.~\ref{fig:Cauchy_saddle-node_1}. Linearizing (\ref{eq:CC_thermo_Lorentzian}) near the stationary point $r^\star = \sqrt{1-2\Delta/K}$, with $r = r^\star + \delta r$ and $|\delta r| \ll 1$, yields
\begin{equation} \nonumber
\dot{\delta r} = \lambda\, \delta r.
\end{equation}
with relaxation rate 
\begin{equation} \label{eq:CC_therm_lim_rel_rate}
\lambda = \epsilon (2\Delta - K).
\end{equation}
It is important to notice that the perturbation $\delta r$ (of the collective coordinate) is a perturbation \textit{along} the ansatz manifold $\hat{\bm{\phi}}$. Note that $\lambda$ converges to zero as $\epsilon \to 0$. This corresponds to the critical slowing down associated with the saddle-node bifurcations shown in Fig.~\ref{fig:Cauchy_saddle-node_1}(b) for finite $N$ and which occurs at every value of $K$ for infinite $N$. It is pertinent to mention that having saddle-node bifurcations for finite networks is consistent with the thermodynamic limit experiencing a pitchfork bifurcation, since pitchfork bifurcations are structurally unstable and transform into a saddle-node bifurcation upon a small perturbation. To understand the transition from a finite network experiencing a cascade of saddle-node bifurcations to an infinite network where saddle-node bifurcations appear at each coupling strength, consider the following approximation of a large but finite network illustrated in Fig.~\ref{fig:rel_rate_cluster}. We consider a continuously distributed frequency distribution on a compact support $[-(Kr - \delta), Kr - \delta]$, for example, the region between the two vertical blue dashed lines in Fig.~\ref{fig:rel_rate_cluster} for $K=4.24$ and $\delta = 1$. The offset $\delta$ denotes the difference between the largest (resp. smallest) natural frequency that will synchronize and $Kr$ (resp. $-Kr$), which, when $\delta \to 0$, yields the set of natural frequencies which will become synchronized. We note that for finite networks, as $N\to \infty$, $\delta$ effectively converges to zero. This is because increased sampling of the frequency distribution $g(\omega)$ yields natural frequencies closer and closer to $Kr$ (or to $-Kr$). We can now write the integral $I_3$ as
\begin{equation}
I_3 = \lim_{\delta \to 0} \tilde{I}_3(\delta), \nonumber
\end{equation}
where 
\begin{align}
\tilde{I}_3(\delta) &= \int_{-(Kr - \delta)}^{Kr - \delta} \left(\frac{d\hat{\phi}}{dr}\right)^2 g(\omega) d\omega \nonumber \\
&= \frac{\Delta}{\pi K r  E^2} \left[ - \log\delta + \log(2Kr - \delta) \vphantom{\arctan\left(\frac{Kr - \delta}{\Delta}\right)} \right. \nonumber \\
&\qquad \qquad \qquad  \left.  - \frac{2\Delta}{Kr} \arctan\left(\frac{Kr - \delta}{\Delta}\right) \right]. \nonumber
\end{align}
The relaxation rate $\lambda$ (\ref{eq:CC_therm_lim_rel_rate}) can analogously be expressed as the limit
\begin{equation}  \label{eq:CC_therm_lim_rel_rate_delta}
\lambda = \lim_{\delta \to 0} \frac{D}{\tilde{I}_3(\delta)}(2\Delta - K).
\end{equation}
Fig.~\ref{fig:rel_rate_vs_delta} shows the relaxation rate $\lambda$ (\ref{eq:CC_therm_lim_rel_rate_delta}) (solid black curve) for $K=4.24$, and, as expected, $\lambda$ converges to zero as $\delta \to 0$ ($-\log(\delta) \to \infty$).

The linearized equation (\ref{eq:CC_therm_lim_rel_rate_delta}) for the order parameter can be connected to the collective coordinate approach for finite networks and to the full Kuramoto model (\ref{eq:full_KM}). The connection to the collective coordinate approach for finite networks is obtained by considering $\tilde{I}_3(\delta)$ as the limit
\begin{equation}
\tilde{I}_3(\delta) = \lim_{N\to \infty} \frac{1}{N} \sum_{i\in \mathcal{C}'(\delta,N)} \left(\frac{d\hat{\bm{\phi}}_i}{d r}\right)^2,  \nonumber
\end{equation}
where 
\begin{equation} \label{eq:finite_CC_delta_cluster}
\mathcal{C}'(\delta,N) = \{i\in \mathcal{C} : \omega_i \in [-(Kr-\delta),Kr-\delta]\}.
\end{equation}
For example, in Fig.~\ref{fig:rel_rate_cluster} the natural frequencies belonging to $\mathcal{C}'(\delta,N)$ for $\delta=1$ and $N=100$ are shown by the blue ticks on the horizontal axis, with the red ticks representing natural frequencies not belonging to $\mathcal{C}'(\delta,N)$. For finite $N$ and a set $\mathcal{C}$, the dynamics along the ansatz manifold is given by (\ref{eq:asin_CC_evolution_equation}), and so the relaxation rate \textit{along} the ansatz manifold toward the stationary state $\hat{\bm{\phi}}(r^\star)$ (i.e., the solution to (\ref{eq:asin_reduced_condition})) is
\begin{align}
\lambda = - \frac{K}{||\frac{d\hat{\bm{\phi}}}{dr}||^2} &\left(\sum_{i \in \mathcal{C}} \frac{s_i^2}{\sqrt{1-s_i^2}} \right) \nonumber  \\
 &  \times\left[ \frac{1}{r^\star} - \frac{1}{N (r^\star)^2} \sum_{j\in \mathcal{C}} \frac{s_j^2}{\sqrt{1-s_j^2}} \right].   \label{eq:finite_CC_relaxation_rate}
\end{align}
This relaxation rate is shown in Fig.~\ref{fig:rel_rate_vs_delta} for $N=10^3$ (green diamonds), $N=10^4$ (orange triangles) and $N=10^5$ (blue circles) with $\mathcal{C} = \mathcal{C}'(\delta,N)$ given by (\ref{eq:finite_CC_delta_cluster}). Fig.~\ref{fig:rel_rate_vs_delta} shows that the relaxation rate (\ref{eq:finite_CC_relaxation_rate}) for finite networks converges to the relaxation rate (\ref{eq:CC_therm_lim_rel_rate_delta}) of the thermodynamic limit as $N\to \infty$ and $\delta \to 0$ ($-\log(\delta) \to \infty$). Hence, the collective coordinate approach in the thermodynamic limit accurately captures the limiting dynamics of the finite population model. In particular, we observe critical slowing down, with $\lambda \to 0$ as $\delta \to 0$, corresponding to the approach to the saddle-node bifurcation. The disagreement between the relaxation rate in the finite and infinite cases for large $\delta$ is due to the fact that in (\ref{eq:CC_thermo_general_form}) only the integral $I_3$ has its support truncated to the interval $[-Kr^\star+\delta,Kr^\star - \delta]$, since the integrals $I_1$ and $I_2$ converge and can be found analytically (cf. (\ref{eq:CC_thermo_Lorentzian})). 

We now provide a geometric interpretation of the linearized collective coordinate equation (\ref{eq:finite_CC_relaxation_rate}). We show that it can be derived directly from the full Kuramoto model (\ref{eq:full_KM}) as the equation describing the relaxation along the ansatz manifold $\hat{\bm{\phi}}(r)$. Recall from (\ref{eq:CC_projection_form}) that the collective coordinate evolution equation can be expressed in the form
\begin{equation} \nonumber
\Pi_{\frac{d\hat{\bm{\phi}}}{dr}} \dot{\bm{\phi}} = \dot{r} \frac{d\hat{\bm{\phi}}}{dr},
\end{equation}
where $\dot{\bm{\phi}}$ is the dynamics of the full Kuramoto model (\ref{eq:full_KM}) and $\Pi_{\frac{d\hat{\bm{\phi}}}{dr}}$ denotes orthogonal projection onto the tangent vector $\frac{d\hat{\bm{\phi}}}{dr}$ (cf. Fig.~\ref{fig:CC_schematic}). Therefore, the relaxation rate obtained from the finite collective coordinate approach (\ref{eq:finite_CC_relaxation_rate}) can be related to the Jacobian of the full Kuramoto model, given by,
\begin{equation}
L(\bm{\phi})_{ij} = \frac{K}{N} \begin{cases}
-\sum_{k\neq i} \cos(\phi_k - \phi_i), &\text{if } i=j  \\
\cos(\phi_j - \phi_i), & \text{if } i\neq	j
\end{cases},
\end{equation}
through the equation
\begin{align}
\lambda &= \frac{ \left\langle \nabla_{\frac{d\hat{\bm{\phi}}}{dr}} \dot{\bm{\phi}}\left( \hat{\bm{\phi}}(r^\star) \right) , \frac{d\hat{\bm{\phi}}}{dr} \right\rangle}
{\left\langle \frac{d\hat{\bm{\phi}}}{dr},\frac{d\hat{\bm{\phi}}}{dr} \right\rangle} \nonumber \\
&=  \frac{ \left\langle L\left(\hat{\bm{\phi}}(r^\star)\right) \frac{d\hat{\bm{\phi}}}{dr} , \frac{d\hat{\bm{\phi}}}{dr} \right\rangle}
{\left\langle \frac{d\hat{\bm{\phi}}}{dr},\frac{d\hat{\bm{\phi}}}{dr} \right\rangle} , \nonumber
\end{align}
where $\nabla_{\bm{v}}$ denotes the directional derivative in the direction $\bm{v}$. It can be readily shown that this reduces to the right hand side of (\ref{eq:finite_CC_relaxation_rate}).

To summarize, the collective coordinate method accurately describes the synchronized states as well as the the dynamics along the ansatz manifold of the full Kuramoto model, for both finite networks and in the thermodynamic limit. The time-scale $\epsilon \to 0$ for the thermodynamic limit reflects the critical slowing down that occurs for finite networks as they approach saddle-node bifurcations that occur at discrete values of the coupling strength $K$ (shown in Fig.~\ref{fig:Cauchy_saddle-node_1}), and that occur in the thermodynamic limit for every value of $K$.

\section{Conclusions} \label{sec:conclusions}

In summary, the collective coordinate framework with both the linear ansatz (\ref{eq:linear_ansatz}) and the arcsin ansatz (\ref{eq:asin_ansatz}) accurately describes the collective dynamics of finite populations of coupled oscillators. The arcsin ansatz yields a significantly improved approximation compared to the linear collective coordinate ansatz. In addition to capturing finite size effects, in the thermodynamic limit the arcsin collective coordinate ansatz recovers well-known analytical results, as well as dynamics of the order parameter along the ansatz manifold.

The collective coordinate approach identifies bifurcations of partially synchronized states as saddle-node bifurcations. When going to the limit $N\to \infty$ we have shown that these saddle-node bifurcations occur at every value of the coupling strength $K$. In the thermodynamic limit, the collective coordinate method predicts a pitchfork bifurcation, consistent with the Ott-Antonsen approach \cite{OttAntonsen08, OttAntonsen09}. The almost continuous cascade of saddle-node bifurcations for finite $N$ is consistent with the thermodynamic limit experiencing a pitchfork bifurcation, since the latter are structurally unstable to any finite size perturbation and transform into a saddle-node bifurcation.

We note that since the $\arcsin$ collective coordinate ansatz (\ref{eq:asin_ansatz}) is based on the mean field formulation of the Kuramoto model (\ref{eq:KM_mean_field}), it is only applicable to networks of globally coupled oscillators. In contrast, the linear collective coordinate ansatz (\ref{eq:linear_ansatz}), which is based on a linearization of the Kuramoto model, can be applied to any network topology \cite{HancockGottwald18,  SmithGottwald19}, and is able to describe partial synchronization in the presence of topological clusters \cite{HancockGottwald18}. However, the arcsin ansatz is well suited to study frequency clustering, and the complex inter- and intra- cluster dynamics that results. Frequency clustering occurs when the natural frequency distribution is multimodal\cite{MartensEtAl09, PazoErnest09, PietrasEtAl18}, or in finite networks when random frequency gaps occur as finite size effects, as well as finite networks with uniformly distributed natural frequencies that exhibit a cascade of frequency splittings \cite{Pazo05}. For such cases, the collective coordinate approach results in a system of coupled equations describing the dynamics of the order parameters and phases of the synchronized clusters \cite{Gottwald15, SmithGottwald19}. The complex non-stationary collective dynamics that results from these systems, including collective chaos \cite{SmithGottwald19}, cannot be described by self-consistency approaches.

Here we have considered only the case of sinusoidal coupling between oscillators. The collective coordinate approach has recently been extended to the Kuramoto-Sakaguchi model which includes a phase-frustration parameter \cite{YueEtAl20}. We believe that the collective coordinate framework can readily be extended to coupling functions with higher harmonics\cite{SkardalEtAl11} and symplectic coupling \cite{SkardalArenas19} by considering multiple synchronized clusters, similar to the case of frequency clustering. This is a topic for future research.

\begin{acknowledgments}
We wish to acknowledge support from the Australian Research Council, Grant No. DP180101991.
\end{acknowledgments}

\section*{Data Availability}

The data that support the findings of this study are available from the corresponding author upon reasonable request.


%\bibliographystyle{alpha}

%\bibliography{Kuramoto_CC_vs_OA.bib}

\begin{thebibliography}{41}%
\makeatletter
\providecommand \@ifxundefined [1]{%
 \@ifx{#1\undefined}
}%
\providecommand \@ifnum [1]{%
 \ifnum #1\expandafter \@firstoftwo
 \else \expandafter \@secondoftwo
 \fi
}%
\providecommand \@ifx [1]{%
 \ifx #1\expandafter \@firstoftwo
 \else \expandafter \@secondoftwo
 \fi
}%
\providecommand \natexlab [1]{#1}%
\providecommand \enquote  [1]{``#1''}%
\providecommand \bibnamefont  [1]{#1}%
\providecommand \bibfnamefont [1]{#1}%
\providecommand \citenamefont [1]{#1}%
\providecommand \href@noop [0]{\@secondoftwo}%
\providecommand \href [0]{\begingroup \@sanitize@url \@href}%
\providecommand \@href[1]{\@@startlink{#1}\@@href}%
\providecommand \@@href[1]{\endgroup#1\@@endlink}%
\providecommand \@sanitize@url [0]{\catcode `\\12\catcode `\$12\catcode
  `\&12\catcode `\#12\catcode `\^12\catcode `\_12\catcode `\%12\relax}%
\providecommand \@@startlink[1]{}%
\providecommand \@@endlink[0]{}%
\providecommand \url  [0]{\begingroup\@sanitize@url \@url }%
\providecommand \@url [1]{\endgroup\@href {#1}{\urlprefix }}%
\providecommand \urlprefix  [0]{URL }%
\providecommand \Eprint [0]{\href }%
\providecommand \doibase [0]{http://dx.doi.org/}%
\providecommand \selectlanguage [0]{\@gobble}%
\providecommand \bibinfo  [0]{\@secondoftwo}%
\providecommand \bibfield  [0]{\@secondoftwo}%
\providecommand \translation [1]{[#1]}%
\providecommand \BibitemOpen [0]{}%
\providecommand \bibitemStop [0]{}%
\providecommand \bibitemNoStop [0]{.\EOS\space}%
\providecommand \EOS [0]{\spacefactor3000\relax}%
\providecommand \BibitemShut  [1]{\csname bibitem#1\endcsname}%
\let\auto@bib@innerbib\@empty
%</preamble>
\bibitem [{\citenamefont {Mirollo}\ and\ \citenamefont
  {Strogatz}(1990)}]{MirolloStrogatz90}%
  \BibitemOpen
  \bibfield  {author} {\bibinfo {author} {\bibfnamefont {R.}~\bibnamefont
  {Mirollo}}\ and\ \bibinfo {author} {\bibfnamefont {S.}~\bibnamefont
  {Strogatz}},\ }\bibfield  {title} {\enquote {\bibinfo {title}
  {Synchronization of pulse-coupled biological oscillators},}\ }\href {\doibase
  10.1137/0150098} {\bibfield  {journal} {\bibinfo  {journal} {SIAM J. Appl.
  Math.}\ }\textbf {\bibinfo {volume} {50}},\ \bibinfo {pages} {1645--1662}
  (\bibinfo {year} {1990})}\BibitemShut {NoStop}%
\bibitem [{\citenamefont {Sheeba}, \citenamefont {Stefanovska},\ and\
  \citenamefont {McClintock}(2008)}]{SheebaEtAl08}%
  \BibitemOpen
  \bibfield  {author} {\bibinfo {author} {\bibfnamefont {J.~H.}\ \bibnamefont
  {Sheeba}}, \bibinfo {author} {\bibfnamefont {A.}~\bibnamefont {Stefanovska}},
  \ and\ \bibinfo {author} {\bibfnamefont {P.~V.~E.}\ \bibnamefont
  {McClintock}},\ }\bibfield  {title} {\enquote {\bibinfo {title} {Neuronal
  synchrony during anesthesia: {A} thalamocortical model},}\ }\href@noop {}
  {\bibfield  {journal} {\bibinfo  {journal} {Biophys. J.}\ }\textbf {\bibinfo
  {volume} {95}},\ \bibinfo {pages} {2722--2727} (\bibinfo {year}
  {2008})}\BibitemShut {NoStop}%
\bibitem [{\citenamefont {Bhowmik}\ and\ \citenamefont
  {Shanahan}(2012)}]{BhowmikShanahan12}%
  \BibitemOpen
  \bibfield  {author} {\bibinfo {author} {\bibfnamefont {D.}~\bibnamefont
  {Bhowmik}}\ and\ \bibinfo {author} {\bibfnamefont {M.}~\bibnamefont
  {Shanahan}},\ }\bibfield  {title} {\enquote {\bibinfo {title} {How well do
  oscillator models capture the behaviour of biological neurons?}}\ }in\
  \href@noop {} {\emph {\bibinfo {booktitle} {The 2012 International Joint
  Conference on Neural Networks (IJCNN)}}}\ (\bibinfo {year} {2012})\ pp.\
  \bibinfo {pages} {1--8}\BibitemShut {NoStop}%
\bibitem [{\citenamefont {Filatrella}, \citenamefont {Nielsen},\ and\
  \citenamefont {Pedersen}(2008)}]{FilatrellaEtAl08}%
  \BibitemOpen
  \bibfield  {author} {\bibinfo {author} {\bibfnamefont {G.}~\bibnamefont
  {Filatrella}}, \bibinfo {author} {\bibfnamefont {A.~H.}\ \bibnamefont
  {Nielsen}}, \ and\ \bibinfo {author} {\bibfnamefont {N.~F.}\ \bibnamefont
  {Pedersen}},\ }\bibfield  {title} {\enquote {\bibinfo {title} {Analysis of a
  power grid using a {K}uramoto-like model},}\ }\href@noop {} {\bibfield
  {journal} {\bibinfo  {journal} {Eur. Phys. J. B}\ }\textbf {\bibinfo {volume}
  {61}},\ \bibinfo {pages} {485--491} (\bibinfo {year} {2008})}\BibitemShut
  {NoStop}%
\bibitem [{\citenamefont {Bick}\ \emph {et~al.}(2020)\citenamefont {Bick},
  \citenamefont {Goodfellow}, \citenamefont {Laing},\ and\ \citenamefont
  {Martens}}]{BickEtAl20}%
  \BibitemOpen
  \bibfield  {author} {\bibinfo {author} {\bibfnamefont {C.}~\bibnamefont
  {Bick}}, \bibinfo {author} {\bibfnamefont {M.}~\bibnamefont {Goodfellow}},
  \bibinfo {author} {\bibfnamefont {C.~R.}\ \bibnamefont {Laing}}, \ and\
  \bibinfo {author} {\bibfnamefont {E.~A.}\ \bibnamefont {Martens}},\
  }\bibfield  {title} {\enquote {\bibinfo {title} {Understanding the dynamics
  of biological and neural oscillator networks through exact mean-field
  reductions: a review},}\ }\href {\doibase 10.1186/s13408-020-00086-9}
  {\bibfield  {journal} {\bibinfo  {journal} {J. Math. Neurosc.}\ }\textbf
  {\bibinfo {volume} {10}},\ \bibinfo {pages} {9} (\bibinfo {year}
  {2020})}\BibitemShut {NoStop}%
\bibitem [{\citenamefont {Kuramoto}(1984)}]{Kuramoto84}%
  \BibitemOpen
  \bibfield  {author} {\bibinfo {author} {\bibfnamefont {Y.}~\bibnamefont
  {Kuramoto}},\ }\href {\doibase 10.1007/978-3-642-69689-3} {\emph {\bibinfo
  {title} {Chemical {O}scillations, {W}aves, and {T}urbulence}}},\ \bibinfo
  {series} {Springer Series in Synergetics}, Vol.~\bibinfo {volume} {19}\
  (\bibinfo  {publisher} {Springer-Verlag},\ \bibinfo {address} {Berlin},\
  \bibinfo {year} {1984})\ pp.\ \bibinfo {pages} {viii+156}\BibitemShut
  {NoStop}%
\bibitem [{\citenamefont {Strogatz}(2000)}]{Strogatz00}%
  \BibitemOpen
  \bibfield  {author} {\bibinfo {author} {\bibfnamefont {S.~H.}\ \bibnamefont
  {Strogatz}},\ }\bibfield  {title} {\enquote {\bibinfo {title} {From
  {K}uramoto to {C}rawford: {E}xploring the onset of synchronization in
  populations of coupled oscillators},}\ }\href {\doibase
  10.1016/S0167-2789(00)00094-4} {\bibfield  {journal} {\bibinfo  {journal}
  {Physica D}\ }\textbf {\bibinfo {volume} {143}},\ \bibinfo {pages} {1--20}
  (\bibinfo {year} {2000})}\BibitemShut {NoStop}%
\bibitem [{\citenamefont {Pikovsky}, \citenamefont {Rosenblum},\ and\
  \citenamefont {Kurths}(2001)}]{PikovskyEtAl01}%
  \BibitemOpen
  \bibfield  {author} {\bibinfo {author} {\bibfnamefont {A.}~\bibnamefont
  {Pikovsky}}, \bibinfo {author} {\bibfnamefont {M.}~\bibnamefont {Rosenblum}},
  \ and\ \bibinfo {author} {\bibfnamefont {J.}~\bibnamefont {Kurths}},\
  }\href@noop {} {\emph {\bibinfo {title} {{Synchronization: {A} {U}niversal
  {C}oncept in {N}onlinear {S}ciences}}}}\ (\bibinfo  {publisher} {Cambridge
  University Press},\ \bibinfo {address} {Cambridge},\ \bibinfo {year}
  {2001})\BibitemShut {NoStop}%
\bibitem [{\citenamefont {Acebr\'on}\ \emph {et~al.}(2005)\citenamefont
  {Acebr\'on}, \citenamefont {Bonilla}, \citenamefont {P\'erez~Vicente},
  \citenamefont {Ritort},\ and\ \citenamefont {Spigler}}]{AcebronEtAl05}%
  \BibitemOpen
  \bibfield  {author} {\bibinfo {author} {\bibfnamefont {J.~A.}\ \bibnamefont
  {Acebr\'on}}, \bibinfo {author} {\bibfnamefont {L.~L.}\ \bibnamefont
  {Bonilla}}, \bibinfo {author} {\bibfnamefont {C.~J.}\ \bibnamefont
  {P\'erez~Vicente}}, \bibinfo {author} {\bibfnamefont {F.}~\bibnamefont
  {Ritort}}, \ and\ \bibinfo {author} {\bibfnamefont {R.}~\bibnamefont
  {Spigler}},\ }\bibfield  {title} {\enquote {\bibinfo {title} {The {K}uramoto
  model: {A} simple paradigm for synchronization phenomena},}\ }\href {\doibase
  10.1103/RevModPhys.77.137} {\bibfield  {journal} {\bibinfo  {journal} {Rev.
  Mod. Phys.}\ }\textbf {\bibinfo {volume} {77}},\ \bibinfo {pages} {137--185}
  (\bibinfo {year} {2005})}\BibitemShut {NoStop}%
\bibitem [{\citenamefont {Osipov}, \citenamefont {Kurths},\ and\ \citenamefont
  {Zhou}(2007)}]{OsipovEtAl07}%
  \BibitemOpen
  \bibfield  {author} {\bibinfo {author} {\bibfnamefont {G.~V.}\ \bibnamefont
  {Osipov}}, \bibinfo {author} {\bibfnamefont {J.}~\bibnamefont {Kurths}}, \
  and\ \bibinfo {author} {\bibfnamefont {C.}~\bibnamefont {Zhou}},\ }\href
  {\doibase 10.1007/978-3-540-71269-5} {\emph {\bibinfo {title}
  {Synchronization in {O}scillatory {N}etworks}}},\ Springer Series in
  Synergetics\ (\bibinfo  {publisher} {Springer},\ \bibinfo {address}
  {Berlin},\ \bibinfo {year} {2007})\ p.\ \bibinfo {pages} {37c}\BibitemShut
  {NoStop}%
\bibitem [{\citenamefont {Arenas}\ \emph {et~al.}(2008)\citenamefont {Arenas},
  \citenamefont {Diaz-Guilera}, \citenamefont {Kurths}, \citenamefont
  {Moreno},\ and\ \citenamefont {Zhou}}]{ArenasEtAl08}%
  \BibitemOpen
  \bibfield  {author} {\bibinfo {author} {\bibfnamefont {A.}~\bibnamefont
  {Arenas}}, \bibinfo {author} {\bibfnamefont {A.}~\bibnamefont
  {Diaz-Guilera}}, \bibinfo {author} {\bibfnamefont {J.}~\bibnamefont
  {Kurths}}, \bibinfo {author} {\bibfnamefont {Y.}~\bibnamefont {Moreno}}, \
  and\ \bibinfo {author} {\bibfnamefont {C.}~\bibnamefont {Zhou}},\ }\bibfield
  {title} {\enquote {\bibinfo {title} {Synchronization in complex networks},}\
  }\href {\doibase 10.1016/j.physrep.2008.09.002} {\bibfield  {journal}
  {\bibinfo  {journal} {Phys. Rep.}\ }\textbf {\bibinfo {volume} {469}},\
  \bibinfo {pages} {93--153} (\bibinfo {year} {2008})}\BibitemShut {NoStop}%
\bibitem [{\citenamefont {D{\"o}rfler}\ and\ \citenamefont
  {Bullo}(2014)}]{DorflerBullo14}%
  \BibitemOpen
  \bibfield  {author} {\bibinfo {author} {\bibfnamefont {F.}~\bibnamefont
  {D{\"o}rfler}}\ and\ \bibinfo {author} {\bibfnamefont {F.}~\bibnamefont
  {Bullo}},\ }\bibfield  {title} {\enquote {\bibinfo {title} {Synchronization
  in complex networks of phase oscillators: {A} survey},}\ }\href@noop {}
  {\bibfield  {journal} {\bibinfo  {journal} {Automatica}\ }\textbf {\bibinfo
  {volume} {50}},\ \bibinfo {pages} {1539 -- 1564} (\bibinfo {year}
  {2014})}\BibitemShut {NoStop}%
\bibitem [{\citenamefont {Rodrigues}\ \emph {et~al.}(2016)\citenamefont
  {Rodrigues}, \citenamefont {Peron}, \citenamefont {Ji},\ and\ \citenamefont
  {Kurths}}]{RodriguesEtAl16}%
  \BibitemOpen
  \bibfield  {author} {\bibinfo {author} {\bibfnamefont {F.~A.}\ \bibnamefont
  {Rodrigues}}, \bibinfo {author} {\bibfnamefont {T.~K.~D.}\ \bibnamefont
  {Peron}}, \bibinfo {author} {\bibfnamefont {P.}~\bibnamefont {Ji}}, \ and\
  \bibinfo {author} {\bibfnamefont {J.}~\bibnamefont {Kurths}},\ }\bibfield
  {title} {\enquote {\bibinfo {title} {The {K}uramoto model in complex
  networks},}\ }\href@noop {} {\bibfield  {journal} {\bibinfo  {journal} {Phys.
  Rep.}\ }\textbf {\bibinfo {volume} {610}},\ \bibinfo {pages} {1 -- 98}
  (\bibinfo {year} {2016})}\BibitemShut {NoStop}%
\bibitem [{\citenamefont {Ott}\ and\ \citenamefont
  {Antonsen}(2008)}]{OttAntonsen08}%
  \BibitemOpen
  \bibfield  {author} {\bibinfo {author} {\bibfnamefont {E.}~\bibnamefont
  {Ott}}\ and\ \bibinfo {author} {\bibfnamefont {T.~M.}\ \bibnamefont
  {Antonsen}},\ }\bibfield  {title} {\enquote {\bibinfo {title} {Low
  dimensional behavior of large systems of globally coupled oscillators},}\
  }\href {\doibase 10.1063/1.2930766} {\bibfield  {journal} {\bibinfo
  {journal} {Chaos}\ }\textbf {\bibinfo {volume} {18}},\ \bibinfo {pages}
  {037113, 6} (\bibinfo {year} {2008})}\BibitemShut {NoStop}%
\bibitem [{\citenamefont {Ott}\ and\ \citenamefont
  {Antonsen}(2009)}]{OttAntonsen09}%
  \BibitemOpen
  \bibfield  {author} {\bibinfo {author} {\bibfnamefont {E.}~\bibnamefont
  {Ott}}\ and\ \bibinfo {author} {\bibfnamefont {T.~M.}\ \bibnamefont
  {Antonsen}},\ }\bibfield  {title} {\enquote {\bibinfo {title} {Long time
  evolution of phase oscillator systems},}\ }\href {\doibase 10.1063/1.3136851}
  {\bibfield  {journal} {\bibinfo  {journal} {Chaos}\ }\textbf {\bibinfo
  {volume} {19}},\ \bibinfo {pages} {023117} (\bibinfo {year}
  {2009})}\BibitemShut {NoStop}%
\bibitem [{\citenamefont {Panaggio}\ and\ \citenamefont
  {Abrams}(2015)}]{PanaggioAbrams15}%
  \BibitemOpen
  \bibfield  {author} {\bibinfo {author} {\bibfnamefont {M.~J.}\ \bibnamefont
  {Panaggio}}\ and\ \bibinfo {author} {\bibfnamefont {D.~M.}\ \bibnamefont
  {Abrams}},\ }\bibfield  {title} {\enquote {\bibinfo {title} {Chimera states:
  coexistence of coherence and incoherence in networks of coupled
  oscillators},}\ }\href {\doibase 10.1088/0951-7715/28/3/r67} {\bibfield
  {journal} {\bibinfo  {journal} {Nonlinearity}\ }\textbf {\bibinfo {volume}
  {28}},\ \bibinfo {pages} {R67--R87} (\bibinfo {year} {2015})}\BibitemShut
  {NoStop}%
\bibitem [{\citenamefont {Laing}(2009{\natexlab{a}})}]{Laing09}%
  \BibitemOpen
  \bibfield  {author} {\bibinfo {author} {\bibfnamefont {C.~R.}\ \bibnamefont
  {Laing}},\ }\bibfield  {title} {\enquote {\bibinfo {title} {Chimera states in
  heterogeneous networks},}\ }\href {\doibase 10.1063/1.3068353} {\bibfield
  {journal} {\bibinfo  {journal} {Chaos}\ }\textbf {\bibinfo {volume} {19}},\
  \bibinfo {pages} {013113} (\bibinfo {year} {2009}{\natexlab{a}})}\BibitemShut
  {NoStop}%
\bibitem [{\citenamefont {Laing}(2009{\natexlab{b}})}]{Laing09_2}%
  \BibitemOpen
  \bibfield  {author} {\bibinfo {author} {\bibfnamefont {C.~R.}\ \bibnamefont
  {Laing}},\ }\bibfield  {title} {\enquote {\bibinfo {title} {The dynamics of
  chimera states in heterogeneous {K}uramoto networks},}\ }\href {\doibase
  https://doi.org/10.1016/j.physd.2009.04.012} {\bibfield  {journal} {\bibinfo
  {journal} {Physica D}\ }\textbf {\bibinfo {volume} {238}},\ \bibinfo {pages}
  {1569 -- 1588} (\bibinfo {year} {2009}{\natexlab{b}})}\BibitemShut {NoStop}%
\bibitem [{\citenamefont {Skardal}, \citenamefont {Ott},\ and\ \citenamefont
  {Restrepo}(2011)}]{SkardalEtAl11}%
  \BibitemOpen
  \bibfield  {author} {\bibinfo {author} {\bibfnamefont {P.~S.}\ \bibnamefont
  {Skardal}}, \bibinfo {author} {\bibfnamefont {E.}~\bibnamefont {Ott}}, \ and\
  \bibinfo {author} {\bibfnamefont {J.~G.}\ \bibnamefont {Restrepo}},\
  }\bibfield  {title} {\enquote {\bibinfo {title} {Cluster synchrony in systems
  of coupled phase oscillators with higher-order coupling},}\ }\href {\doibase
  10.1103/PhysRevE.84.036208} {\bibfield  {journal} {\bibinfo  {journal} {Phys.
  Rev. E}\ }\textbf {\bibinfo {volume} {84}},\ \bibinfo {pages} {036208}
  (\bibinfo {year} {2011})}\BibitemShut {NoStop}%
\bibitem [{\citenamefont {Skardal}\ and\ \citenamefont
  {Arenas}(2019)}]{SkardalArenas19}%
  \BibitemOpen
  \bibfield  {author} {\bibinfo {author} {\bibfnamefont {P.~S.}\ \bibnamefont
  {Skardal}}\ and\ \bibinfo {author} {\bibfnamefont {A.}~\bibnamefont
  {Arenas}},\ }\bibfield  {title} {\enquote {\bibinfo {title} {Abrupt
  desynchronization and extensive multistability in globally coupled oscillator
  simplexes},}\ }\href {\doibase 10.1103/PhysRevLett.122.248301} {\bibfield
  {journal} {\bibinfo  {journal} {Phys. Rev. Lett.}\ }\textbf {\bibinfo
  {volume} {122}},\ \bibinfo {pages} {248301} (\bibinfo {year}
  {2019})}\BibitemShut {NoStop}%
\bibitem [{\citenamefont {Bick}, \citenamefont {Panaggio},\ and\ \citenamefont
  {Martens}(2018)}]{BickEtAl18}%
  \BibitemOpen
  \bibfield  {author} {\bibinfo {author} {\bibfnamefont {C.}~\bibnamefont
  {Bick}}, \bibinfo {author} {\bibfnamefont {M.~J.}\ \bibnamefont {Panaggio}},
  \ and\ \bibinfo {author} {\bibfnamefont {E.~A.}\ \bibnamefont {Martens}},\
  }\bibfield  {title} {\enquote {\bibinfo {title} {Chaos in {K}uramoto
  oscillator networks},}\ }\href {\doibase 10.1063/1.5041444} {\bibfield
  {journal} {\bibinfo  {journal} {Chaos}\ }\textbf {\bibinfo {volume} {28}},\
  \bibinfo {pages} {071102} (\bibinfo {year} {2018})}\BibitemShut {NoStop}%
\bibitem [{\citenamefont {Paz\'o}\ and\ \citenamefont
  {Montbri\'o}(2009)}]{PazoErnest09}%
  \BibitemOpen
  \bibfield  {author} {\bibinfo {author} {\bibfnamefont {D.}~\bibnamefont
  {Paz\'o}}\ and\ \bibinfo {author} {\bibfnamefont {E.}~\bibnamefont
  {Montbri\'o}},\ }\bibfield  {title} {\enquote {\bibinfo {title} {Existence of
  hysteresis in the {K}uramoto model with bimodal frequency distributions},}\
  }\href {\doibase 10.1103/PhysRevE.80.046215} {\bibfield  {journal} {\bibinfo
  {journal} {Phys. Rev. E}\ }\textbf {\bibinfo {volume} {80}},\ \bibinfo
  {pages} {046215} (\bibinfo {year} {2009})}\BibitemShut {NoStop}%
\bibitem [{\citenamefont {Lu\c{c}on}(2015)}]{Lucon15}%
  \BibitemOpen
  \bibfield  {author} {\bibinfo {author} {\bibfnamefont {E.}~\bibnamefont
  {Lu\c{c}on}},\ }\bibfield  {title} {\enquote {\bibinfo {title} {Large
  population asymptotics for interacting diffusions in a quenched random
  environment},}\ }in\ \href@noop {} {\emph {\bibinfo {booktitle} {From
  particle systems to partial differential equations. {II}}}},\ \bibinfo
  {series} {Springer Proc. Math. Stat.}, Vol.\ \bibinfo {volume} {129}\
  (\bibinfo  {publisher} {Springer, Cham},\ \bibinfo {year} {2015})\ pp.\
  \bibinfo {pages} {231--251}\BibitemShut {NoStop}%
\bibitem [{\citenamefont {Bertini}, \citenamefont {Giacomin},\ and\
  \citenamefont {Pakdaman}(2010)}]{BertiniEtAl10}%
  \BibitemOpen
  \bibfield  {author} {\bibinfo {author} {\bibfnamefont {L.}~\bibnamefont
  {Bertini}}, \bibinfo {author} {\bibfnamefont {G.}~\bibnamefont {Giacomin}}, \
  and\ \bibinfo {author} {\bibfnamefont {K.}~\bibnamefont {Pakdaman}},\
  }\bibfield  {title} {\enquote {\bibinfo {title} {Dynamical aspects of mean
  field plane rotators and the {K}uramoto model},}\ }\href@noop {} {\bibfield
  {journal} {\bibinfo  {journal} {J. Stat. Phys.}\ }\textbf {\bibinfo {volume}
  {138}},\ \bibinfo {pages} {270--290} (\bibinfo {year} {2010})}\BibitemShut
  {NoStop}%
\bibitem [{\citenamefont {Bertini}, \citenamefont {Giacomin},\ and\
  \citenamefont {Poquet}(2014)}]{BertiniEtAl14}%
  \BibitemOpen
  \bibfield  {author} {\bibinfo {author} {\bibfnamefont {L.}~\bibnamefont
  {Bertini}}, \bibinfo {author} {\bibfnamefont {G.}~\bibnamefont {Giacomin}}, \
  and\ \bibinfo {author} {\bibfnamefont {C.}~\bibnamefont {Poquet}},\
  }\bibfield  {title} {\enquote {\bibinfo {title} {Synchronization and random
  long time dynamics for mean-field plane rotators},}\ }\href@noop {}
  {\bibfield  {journal} {\bibinfo  {journal} {Probab. Theory Related Fields}\
  }\textbf {\bibinfo {volume} {160}},\ \bibinfo {pages} {593--653} (\bibinfo
  {year} {2014})}\BibitemShut {NoStop}%
\bibitem [{\citenamefont {Gottwald}(2017)}]{Gottwald17}%
  \BibitemOpen
  \bibfield  {author} {\bibinfo {author} {\bibfnamefont {G.~A.}\ \bibnamefont
  {Gottwald}},\ }\bibfield  {title} {\enquote {\bibinfo {title} {Finite-size
  effects in a stochastic {K}uramoto model},}\ }\href {\doibase
  10.1063/1.5004618} {\bibfield  {journal} {\bibinfo  {journal} {Chaos}\
  }\textbf {\bibinfo {volume} {27}},\ \bibinfo {pages} {101103} (\bibinfo
  {year} {2017})}\BibitemShut {NoStop}%
\bibitem [{\citenamefont {Watanabe}\ and\ \citenamefont
  {Strogatz}(1993)}]{WatanabeStrogatz93}%
  \BibitemOpen
  \bibfield  {author} {\bibinfo {author} {\bibfnamefont {S.}~\bibnamefont
  {Watanabe}}\ and\ \bibinfo {author} {\bibfnamefont {S.~H.}\ \bibnamefont
  {Strogatz}},\ }\bibfield  {title} {\enquote {\bibinfo {title} {Integrability
  of a globally coupled oscillator array},}\ }\href@noop {} {\bibfield
  {journal} {\bibinfo  {journal} {Phys. Rev. Lett.}\ }\textbf {\bibinfo
  {volume} {70}},\ \bibinfo {pages} {2391--2394} (\bibinfo {year}
  {1993})}\BibitemShut {NoStop}%
\bibitem [{\citenamefont {Pikovsky}\ and\ \citenamefont
  {Rosenblum}(2015)}]{RosenblumPikovsky15}%
  \BibitemOpen
  \bibfield  {author} {\bibinfo {author} {\bibfnamefont {A.}~\bibnamefont
  {Pikovsky}}\ and\ \bibinfo {author} {\bibfnamefont {M.}~\bibnamefont
  {Rosenblum}},\ }\bibfield  {title} {\enquote {\bibinfo {title} {Dynamics of
  globally coupled oscillators: {P}rogress and perspectives},}\ }\href@noop {}
  {\bibfield  {journal} {\bibinfo  {journal} {Chaos}\ }\textbf {\bibinfo
  {volume} {25}},\ \bibinfo {pages} {097616} (\bibinfo {year}
  {2015})}\BibitemShut {NoStop}%
\bibitem [{\citenamefont {Pikovsky}\ and\ \citenamefont
  {Rosenblum}(2011)}]{PikovskyRosenblum11}%
  \BibitemOpen
  \bibfield  {author} {\bibinfo {author} {\bibfnamefont {A.}~\bibnamefont
  {Pikovsky}}\ and\ \bibinfo {author} {\bibfnamefont {M.}~\bibnamefont
  {Rosenblum}},\ }\bibfield  {title} {\enquote {\bibinfo {title} {Dynamics of
  heterogeneous oscillator ensembles in terms of collective variables},}\
  }\href {\doibase https://doi.org/10.1016/j.physd.2011.01.002} {\bibfield
  {journal} {\bibinfo  {journal} {Physica D}\ }\textbf {\bibinfo {volume}
  {240}},\ \bibinfo {pages} {872 -- 881} (\bibinfo {year} {2011})}\BibitemShut
  {NoStop}%
\bibitem [{\citenamefont {Gottwald}(2015)}]{Gottwald15}%
  \BibitemOpen
  \bibfield  {author} {\bibinfo {author} {\bibfnamefont {G.~A.}\ \bibnamefont
  {Gottwald}},\ }\bibfield  {title} {\enquote {\bibinfo {title} {Model
  reduction for networks of coupled oscillators},}\ }\href@noop {} {\bibfield
  {journal} {\bibinfo  {journal} {Chaos}\ }\textbf {\bibinfo {volume} {25}},\
  \bibinfo {pages} {053111, 12} (\bibinfo {year} {2015})}\BibitemShut {NoStop}%
\bibitem [{\citenamefont {Hancock}\ and\ \citenamefont
  {Gottwald}(2018)}]{HancockGottwald18}%
  \BibitemOpen
  \bibfield  {author} {\bibinfo {author} {\bibfnamefont {E.~J.}\ \bibnamefont
  {Hancock}}\ and\ \bibinfo {author} {\bibfnamefont {G.~A.}\ \bibnamefont
  {Gottwald}},\ }\bibfield  {title} {\enquote {\bibinfo {title} {Model
  reduction for {K}uramoto models with complex topologies},}\ }\href {\doibase
  10.1103/PhysRevE.98.012307} {\bibfield  {journal} {\bibinfo  {journal} {Phys.
  Rev. E}\ }\textbf {\bibinfo {volume} {98}},\ \bibinfo {pages} {012307}
  (\bibinfo {year} {2018})}\BibitemShut {NoStop}%
\bibitem [{\citenamefont {Smith}\ and\ \citenamefont
  {Gottwald}(2019)}]{SmithGottwald19}%
  \BibitemOpen
  \bibfield  {author} {\bibinfo {author} {\bibfnamefont {L.~D.}\ \bibnamefont
  {Smith}}\ and\ \bibinfo {author} {\bibfnamefont {G.~A.}\ \bibnamefont
  {Gottwald}},\ }\bibfield  {title} {\enquote {\bibinfo {title} {Chaos in
  networks of coupled oscillators with multimodal natural frequency
  distributions},}\ }\href {\doibase 10.1063/1.5109130} {\bibfield  {journal}
  {\bibinfo  {journal} {Chaos}\ }\textbf {\bibinfo {volume} {29}},\ \bibinfo
  {pages} {093127} (\bibinfo {year} {2019})}\BibitemShut {NoStop}%
\bibitem [{\citenamefont {Yue}, \citenamefont {Smith},\ and\ \citenamefont
  {Gottwald}(2020)}]{YueEtAl20}%
  \BibitemOpen
  \bibfield  {author} {\bibinfo {author} {\bibfnamefont {W.}~\bibnamefont
  {Yue}}, \bibinfo {author} {\bibfnamefont {L.~D.}\ \bibnamefont {Smith}}, \
  and\ \bibinfo {author} {\bibfnamefont {G.~A.}\ \bibnamefont {Gottwald}},\
  }\bibfield  {title} {\enquote {\bibinfo {title} {Model reduction for the
  kuramoto-sakaguchi model: The importance of nonentrained rogue
  oscillators},}\ }\href {\doibase 10.1103/PhysRevE.101.062213} {\bibfield
  {journal} {\bibinfo  {journal} {Phys. Rev. E}\ }\textbf {\bibinfo {volume}
  {101}},\ \bibinfo {pages} {062213} (\bibinfo {year} {2020})}\BibitemShut
  {NoStop}%
\bibitem [{Note1()}]{Note1}%
  \BibitemOpen
  \bibinfo {note} {This is a standard procedure \cite {Kuramoto84, Strogatz00}.
  The coordinate transformation is $\phi (t) \to \phi (t) - \Omega t - \psi $,
  where $\Omega $ is the mean frequency of the cluster.}\BibitemShut {Stop}%
\bibitem [{\citenamefont {Mirollo}\ and\ \citenamefont
  {Strogatz}(2005)}]{MirolloStrogatz05}%
  \BibitemOpen
  \bibfield  {author} {\bibinfo {author} {\bibfnamefont {R.~E.}\ \bibnamefont
  {Mirollo}}\ and\ \bibinfo {author} {\bibfnamefont {S.~H.}\ \bibnamefont
  {Strogatz}},\ }\bibfield  {title} {\enquote {\bibinfo {title} {The spectrum
  of the locked state for the {K}uramoto model of coupled oscillators},}\
  }\href {\doibase https://doi.org/10.1016/j.physd.2005.01.017} {\bibfield
  {journal} {\bibinfo  {journal} {Physica D}\ }\textbf {\bibinfo {volume}
  {205}},\ \bibinfo {pages} {249 -- 266} (\bibinfo {year} {2005})},\ \bibinfo
  {note} {synchronization and Pattern Formation in Nonlinear Systems: New
  Developments and Future Perspectives}\BibitemShut {NoStop}%
\bibitem [{\citenamefont {Martens}\ \emph {et~al.}(2009)\citenamefont
  {Martens}, \citenamefont {Barreto}, \citenamefont {Strogatz}, \citenamefont
  {Ott}, \citenamefont {So},\ and\ \citenamefont {Antonsen}}]{MartensEtAl09}%
  \BibitemOpen
  \bibfield  {author} {\bibinfo {author} {\bibfnamefont {E.~A.}\ \bibnamefont
  {Martens}}, \bibinfo {author} {\bibfnamefont {E.}~\bibnamefont {Barreto}},
  \bibinfo {author} {\bibfnamefont {S.~H.}\ \bibnamefont {Strogatz}}, \bibinfo
  {author} {\bibfnamefont {E.}~\bibnamefont {Ott}}, \bibinfo {author}
  {\bibfnamefont {P.}~\bibnamefont {So}}, \ and\ \bibinfo {author}
  {\bibfnamefont {T.~M.}\ \bibnamefont {Antonsen}},\ }\bibfield  {title}
  {\enquote {\bibinfo {title} {Exact results for the {K}uramoto model with a
  bimodal frequency distribution},}\ }\href {\doibase
  10.1103/PhysRevE.79.026204} {\bibfield  {journal} {\bibinfo  {journal} {Phys.
  Rev. E}\ }\textbf {\bibinfo {volume} {79}},\ \bibinfo {pages} {026204}
  (\bibinfo {year} {2009})}\BibitemShut {NoStop}%
\bibitem [{\citenamefont {Pietras}, \citenamefont {Deschle},\ and\
  \citenamefont {Daffertshofer}(2018)}]{PietrasEtAl18}%
  \BibitemOpen
  \bibfield  {author} {\bibinfo {author} {\bibfnamefont {B.}~\bibnamefont
  {Pietras}}, \bibinfo {author} {\bibfnamefont {N.}~\bibnamefont {Deschle}}, \
  and\ \bibinfo {author} {\bibfnamefont {A.}~\bibnamefont {Daffertshofer}},\
  }\bibfield  {title} {\enquote {\bibinfo {title} {First-order phase
  transitions in the {K}uramoto model with compact bimodal frequency
  distributions},}\ }\href {\doibase 10.1103/PhysRevE.98.062219} {\bibfield
  {journal} {\bibinfo  {journal} {Phys. Rev. E}\ }\textbf {\bibinfo {volume}
  {98}},\ \bibinfo {pages} {062219} (\bibinfo {year} {2018})}\BibitemShut
  {NoStop}%
\bibitem [{\citenamefont {Omel'chenko}\ and\ \citenamefont
  {Wolfrum}(2012)}]{OmelchenkoWolfrum12}%
  \BibitemOpen
  \bibfield  {author} {\bibinfo {author} {\bibfnamefont {O.~E.}\ \bibnamefont
  {Omel'chenko}}\ and\ \bibinfo {author} {\bibfnamefont {M.}~\bibnamefont
  {Wolfrum}},\ }\bibfield  {title} {\enquote {\bibinfo {title} {{Nonuniversal
  Transitions to Synchrony in the Sakaguchi-Kuramoto Model}},}\ }\href
  {\doibase 10.1103/PhysRevLett.109.164101} {\bibfield  {journal} {\bibinfo
  {journal} {Phys. Rev. Lett.}\ }\textbf {\bibinfo {volume} {109}},\ \bibinfo
  {pages} {164101} (\bibinfo {year} {2012})}\BibitemShut {NoStop}%
\bibitem [{\citenamefont {Omel’chenko}\ and\ \citenamefont
  {Wolfrum}(2013)}]{OmelchenkoWolfrum13}%
  \BibitemOpen
  \bibfield  {author} {\bibinfo {author} {\bibfnamefont {O.~E.}\ \bibnamefont
  {Omel’chenko}}\ and\ \bibinfo {author} {\bibfnamefont {M.}~\bibnamefont
  {Wolfrum}},\ }\bibfield  {title} {\enquote {\bibinfo {title} {Bifurcations in
  the {S}akaguchi–{K}uramoto model},}\ }\href {\doibase
  https://doi.org/10.1016/j.physd.2013.08.004} {\bibfield  {journal} {\bibinfo
  {journal} {Physica D}\ }\textbf {\bibinfo {volume} {263}},\ \bibinfo {pages}
  {74 -- 85} (\bibinfo {year} {2013})}\BibitemShut {NoStop}%
\bibitem [{\citenamefont {Paz{\'o}}(2005)}]{Pazo05}%
  \BibitemOpen
  \bibfield  {author} {\bibinfo {author} {\bibfnamefont {D.}~\bibnamefont
  {Paz{\'o}}},\ }\bibfield  {title} {\enquote {\bibinfo {title} {Thermodynamic
  limit of the first-order phase transition in the {K}uramoto model},}\ }\href
  {\doibase 10.1103/PhysRevE.72.046211} {\bibfield  {journal} {\bibinfo
  {journal} {Phys. Rev. E}\ }\textbf {\bibinfo {volume} {72}},\ \bibinfo
  {pages} {046211, 6} (\bibinfo {year} {2005})}\BibitemShut {NoStop}%
\bibitem [{\citenamefont {Skardal}(2018)}]{Skardal18}%
  \BibitemOpen
  \bibfield  {author} {\bibinfo {author} {\bibfnamefont {P.~S.}\ \bibnamefont
  {Skardal}},\ }\bibfield  {title} {\enquote {\bibinfo {title} {Low-dimensional
  dynamics of the {K}uramoto model with rational frequency distributions},}\
  }\href {\doibase 10.1103/PhysRevE.98.022207} {\bibfield  {journal} {\bibinfo
  {journal} {Phys. Rev. E}\ }\textbf {\bibinfo {volume} {98}},\ \bibinfo
  {pages} {022207} (\bibinfo {year} {2018})}\BibitemShut {NoStop}%
\end{thebibliography}

\end{document}